\documentclass[useAMS,usenatbib]{mn2e}

\usepackage{amssymb}
\usepackage{times}
\usepackage{graphicx}

\title[Cross-correlation Study between the Cosmological 21-cm Signal
and the kinetic Sunyaev-Zel'dovich effect]{Cross-correlation Study
between the Cosmological 21-cm Signal and the Kinetic
Sunyaev-Zel'dovich Effect}

\author[Jeli{\'c} et al. ]{Vibor Jeli{\'c}$^{1}$\thanks{E-mail:
vjelic@astro.rug.nl}, Saleem Zaroubi$^{1}$, Nabila Aghanim$^{2}$,
Marian Douspis$^{2}$, L\'{e}on V. E. Koopmans$^{1}$, \newauthor
Mathieu Langer$^{2}$, Garrelt Mellema$^{3}$, Hiroyuki Tashiro$^{2}$,
Rajat M. Thomas$^{1,4}$ \\ $^{1}$Kapteyn Astronomical Institute,
University of Groningen, P.O. Box 800, 9700 AV Groningen, the
Netherlands\\ $^{2}$Institut d'Astrophysique Spatiale (IAS),
B{\^a}timent 121, F-91405, Orsay, France; Universit{\'e} Paris-Sud XI
and CNRS (UMR 8617), France\\ $^3$Stockholm Observatory, AlbaNova
University Center, Stockholm University, SE-106 91, Stockholm, Sweden
\\ $^4$Institute for the Mathematics and Physics of the Universe
(IPMU), The University of Tokyo, Chiba 277-8582, Japan}

\begin{document}


\pagerange{\pageref{firstpage}--\pageref{lastpage}} \pubyear{2009}

\maketitle

\label{firstpage}

\begin{abstract}
The Universe's Epoch of Reionization (EoR) can be studied using a
number of observational probes that provide complementary or
corroborating information. Each of these probes suffers from its own
systematic and statistical uncertainties. It is therefore useful to
consider the mutual information that these data sets contain. In this
paper we present a cross-correlation study between the kinetic
Sunyaev-Zel'dovich effect (kSZ) -- produced by the scattering of CMB
photons off free electrons produced during the reionization process --
and the cosmological 21~cm signal -- which reflects the neutral
hydrogen content of the Universe, as a function of redshift.  The
study is carried out using a simulated reionization history in
$100~h^{-1}~{\rm Mpc}$ scale N-body simulations with radiative
transfer. In essence we find that the two probes anti-correlate. The
significance of the anti-correlation signal depends on the extent of
the reionization process, wherein extended histories result in a much
stronger signal compared to instantaneous cases.  Unfortunately
however, once the primary CMB fluctuations are included into our
simulation they serve as a source of large correlated noise that renders
the cross-correlation signal insignificant, regardless of the
reionization scenario.
\end{abstract}


\section{Introduction}
The Epoch of Reionization (EoR) is one of the least explored periods
in the history of the Universe. At present, there are only a
few tentative observational constrains on the EoR like the
Gunn-Peterson troughs \citep{gunn65, fan06} and the cosmic microwave
background (CMB) E-mode polarization \citep{page07} at large
scales. Both of these observations provide strong yet limited
constraints on the EoR. In the near future, however, a number of
observations at various wavelengths (e.g., redshifted 21~cm from HI,
Lyman-$\alpha$ emitters, high redshift QSOs, etc.) are expected to
probe this pivotal epoch in much greater detail. Among these, the
cosmological 21~cm transition line of neutral hydrogen is the most
promising probe of the intergalactic medium (IGM) during reionization
\citep{madau97}

A number of radio telescopes
(e.g. LOFAR\footnote{http://www.lofar.org},
MWA\footnote{http://www.haystack.mit.edu/ast/arrays/mwa}, and
SKA\footnote{http://www.skatelescope.org}) are currently being
constructed/designed that aim at detecting the redshifted 21~cm line
to study the EoR. Unfortunately, these experiments will suffer from a
high degree of contamination, due both to astrophysical interlopers
like the Galactic and extra-galactic foregrounds and non-astrophysical
instrumental effects \citep[e.g.][]{jelic08,labrop09}. Fortunately,
the signal has some characteristics which differentiates it from the
foregrounds and noise, and using proper statistics may extract
signatures of reionization \citep[e.g.][]{furlanetto04, harker09a,
harker09b}. In order to reliably detect the cosmological signal from
the observed data, it is essential to understand in detail all aspects
of the data and their influence on the extracted signal.

Given the challenges and uncertainities involved in measuring the
redshifted 21-cm signal from the EoR, it is vital to corroborate this
result with other probes of the EoR.  In this paper we study the
information imprinted on the CMB by the EoR and its cross~correlation
with the 21~cm probe. Given the launch of the recent PLANCK satellite,
that will measure the CMB with unprecented accuracy, it is only
fitting to conduct a rigorous study into the cross-correlation of
these data sets.

One of the leading sources of secondary anisotropy in the CMB is due
to the scattering of CMB photons off free electrons, created during
the reionization process, along the line of sight (LOS)
\citep{zeldovich69}. These anisotropies when induced by thermal
motions of free electrons is called the thermal Sunyaev-Zel'dovich
effect (tSZ) and whence due to bulk motion of free electrons, the
kinetic Sunyaev-Zel'dovich effect (kSZ). The latter is far more
dominant during reionization \citep[for a review of secondary CMB
effects see][]{aghanim08}.

The kSZ effect from a homogeneously ionized medium, i.e., with ionized
fraction only a function of redshift, has been studied both
analytically and numerically by a number of authors; the linear regime
of this effect first calculated by \citet{sunyaev70} and subsequently
revisited by \citet{ostriker86} and \citet{vishniac87} -- hence also
referred to as the Ostriker-Vishniac (OV) effect. In recent years
various groups have calculated this effect in its non-linear regime
using semi-analytical models and numerical simulations
\citep{gnedin01, santos03, zhang04}. These studies show the
contribution due to non-linear effects being important only at small
angular scales ($l>1000$), while the OV effect dominate at large
angular scales.

The kSZ effect from patchy reionization was first estimated using
simplified semi-analytical models \citep{santos03} wherein they
concluded that fluctuations caused by patchy reionization dominate
over anisotropies induced by homogeneous reionization. However, for a
complete picture of the CMB anisotropies induced by the EoR a more
detailed modeling is required. Over and above the underlying density
and velocity fields these details should include the formation history
and ``nature'' of the first ionizing sources and the radiative
transport of ionizing photons to derive the reionization history
(sizes and distribution of the ionized bubbles). Some recent numerical
simulations of the kSZ effect during the EoR were carried out by
\citet{salvaterra05,zahn05, dore07, iliev07}.

Cross-correlation between the cosmological 21~cm signal and the
secondary CMB anisotropies provide a potentially useful statistic. The
cross-correlation has the advantage that the measured statistic is
less sensitive to contaminants such as the foregrounds, systematics
and noise, in comparison to ``auto-correlation'' studies.  Analytical
cross-correlation studies between the CMB temperature anisotropies and
the EoR signal on large scales ($l\sim100$) have been carried out by
\citet{alvarez06, adshead08} and on small scales ($l>1000$) by
\citet{cooray04, salvaterra05, slosar07}. Thus far the only numerical
study of the cross-correlation has been carried out by
\citet{salvaterra05}. Some additional analytical work on
cross-correlation between the E- and B-modes of CMB polarization with
the redshifted 21~cm signal have been done by \cite{tashiro08,
dvorkin09}.

In this paper we first calculate the kSZ anisotropies from the
homogeneous and patchy reionization based on $100~h^{-1}~{\rm Mpc}$
scale numerical simulations of reionization. We then cross-correlate
them with the expected EoR maps obtained via the same simulations, and
we discuss how the large-scale velocites and primary CMB fluctuations
influence the cross-correlation. Although similar in some aspects, the
work presented here differ from \citet{salvaterra05}
substantially. First, Salvaterra et al. used a relatively small
computational box ($20~h^{-1}~{\rm Mpc}$) incapable of capturing
relevant large-scale density and velocity pertubations. Secondly, the
primary CMB fluctuation, that manifests itself as a large background
noise, is not taken into the cross-correlation study. And finally,
there is a difference in procedure for calculating that
cross-correlation coefficient.

The paper is organized as follows. In Section~\ref{sec:theory} we
discuss kSZ signal and cosmological 21~cm signal from the EoR. In
Section~\ref{sec:sim} we present the numerical simulations employed to
obtain the kSZ and EoR maps for a specifc reionization
history. Cross-correlation between the cosmological 21~cm fluctuations
(EoR signal) and the kSZ anisotropies, together with the influence of
the large-scale velocities and the primary CMB fluctuations on CMB-EoR
cross-correlation are discussed in Section~\ref{sec:cross}. Finally in
Section~\ref{sec:dc} we present our discussions and conclusions on the
topic.

Throughout we assume $\Lambda$CDM-cosmology with WMAP5 parameters
\citep{komatsu08}: $H_0=71.9~{\rm kms^{-1}Mpc^{-1}}$, $\Omega_{\rm b}=0.0441$,
$\Omega_{\rm m}=0.258$ and $\Omega_\Lambda=0.742$.

\section{Theory}\label{sec:theory}
In the following section we briefly review the theoretical aspects of
the kinetic Sunayev-Zel'dovich (kSZ) effect and the cosmological 21~cm
signal from the epoch of reionization. In addition, the relevant
mathematical forms used to calculate the kSZ and the cosmological
21~cm signals are presented.

\subsection{Kinetic Sunayev-Zel'dovich effect}
\label{sec:kSZ}

\begin{figure*}
\centering \includegraphics[width=.95\textwidth]{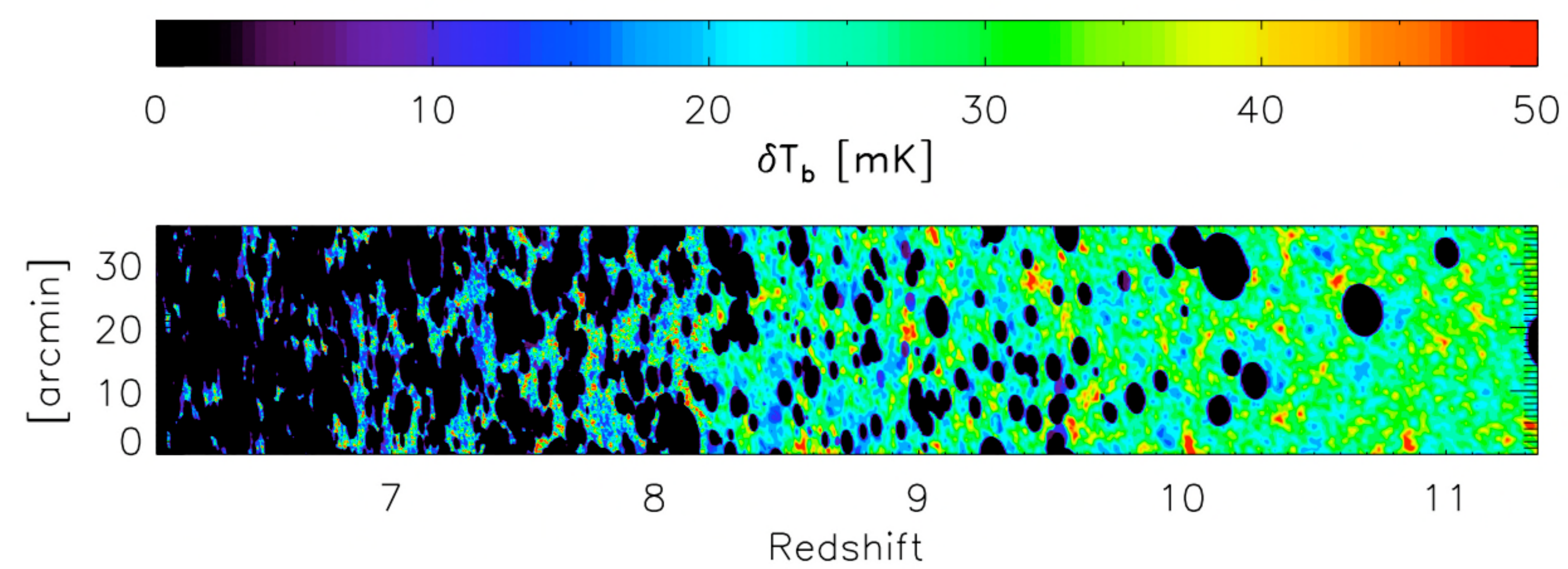}
\centering \includegraphics[width=.95\textwidth]{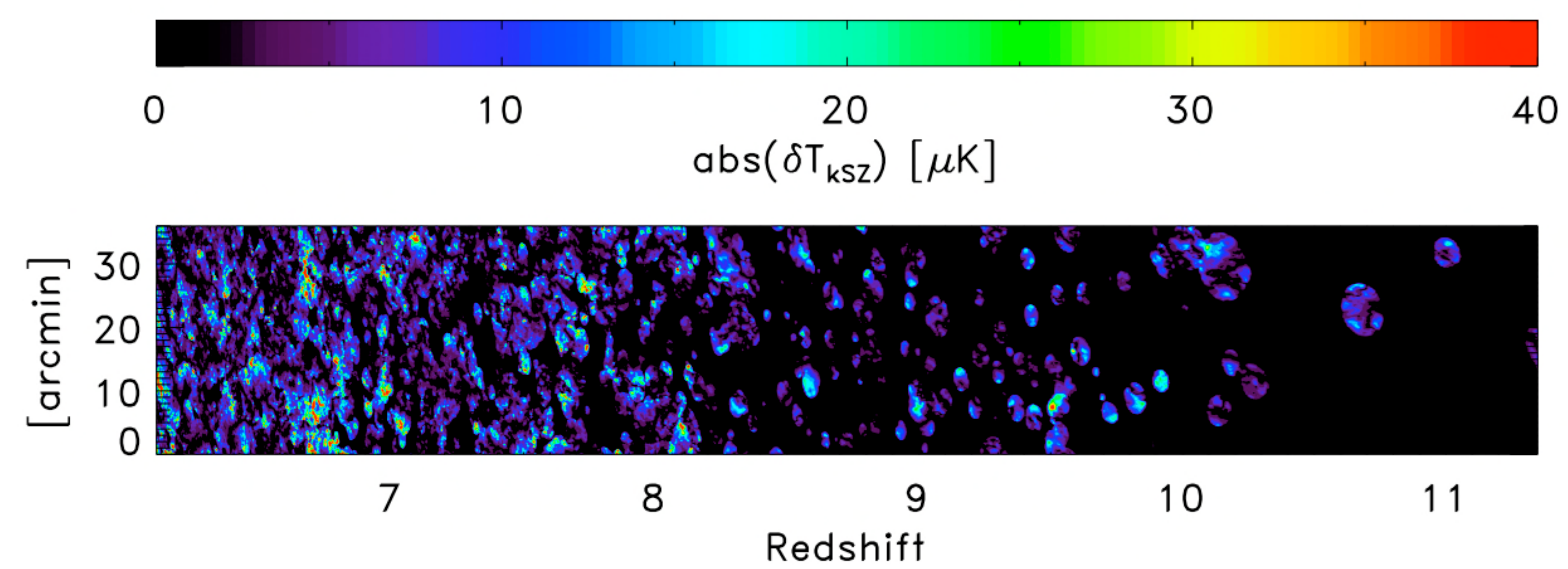}
\caption{\emph{A slice through the simulated redshift cube of the
the cosmological 21~cm signal (top panel) and the kSZ effect (bottom
panel) in the case of the `Stars' patchy reionization model. The
angular scale of the slices is $\sim 0.6^\circ$.}}
\label{fig:EoRkSZs}
\end{figure*}

The temperature fluctuation of the CMB caused by the Thompson
scattering of its photons off populations of free electrons in
bulk motion, for a given line of sight (LOS), is:

\begin{equation}\label{eq:dTkCMB}
\left( \frac{\delta T}{T} \right)_{\rm kSZ} = - {\rm \sigma_{T}}
\int_{t_{\rm r}}^{t_{0}} {\rm e}^{-\tau} n_{\rm e} (\hat{r} \cdot
\vec{v}) \mathrm{d} t,
\end{equation}
where $\tau$ is the optical depth of electrons to Thomson scattering,
$\vec{v}$ the bulk velocity of free electrons and $\hat{r}$ the unit
vector denoting the direction of the LOS. The integral is
performed for each LOS with $t_{\rm r}$  the time at the
epoch recombination and $t_{0}$ the age of the Universe today. Note
that all quantities are in physical units. Temperature fluctuations
produced at time $t$ will be attenuated due to multiple scattering
along the LOS to the present time and is accounted for by
the ${\rm e}^{-\tau}$ term.

The electron density can be written as the product of the neutral atom
density $n_{\rm n}$ and ionization fraction $x_{\rm e}$. Both $n_{\rm
n}$ and $x_{\rm e}$ vary about their average value $\bar{n}_{\rm n}$
and $\bar{x_{\rm e}}$ and thus these fluctuations can be defined as
$\delta=n_{\rm n}/\bar{n}_{\rm n}-1$ and $\delta_{x_{\rm e}}=x_{\rm
e}/\bar{x}_{\rm e}-1$ and consequently the electron density expressed
as:

\begin{equation}\label{eq:ne}
n_{\rm e}=\bar{n}_{\rm n}\bar{x}_{\rm e}(1 + \delta + \delta_{x_{\rm e}}
+ \delta\delta_{x_{\rm e}}).
\end{equation}

In first approximation one can just follow the reionization of
hydrogen and assume that the neutral atom density equals neutral
hydrogen density. However, in our simulation we follow both 
hydrogen and helium reionization. Assuming both hydrogen and helium
follow the underlying dark matter density, the neutral atom density is
a sum of the neutral hydrogen ($n_{\rm HI}$) and neutral helium
($n_{\rm HeI}$) densities: $n_{\rm n}=(\bar{n}_{\rm HI}+\bar{n}_{\rm
HeI})(1+\delta)$. Moreover, the electron density can be written as:
\begin{equation}\label{eq:ne2}
n_{\rm e}=n_{\rm HI}x_{\rm HII}+n_{\rm HeI}x_{\rm HeII}+2n_{\rm
HeI}x_{\rm HeIII},
\end{equation}
where $n_{\rm HII, HeII, HeIII}$ and $x_{\rm HII, HeII, HeIII}$ are
densities and ionization fractions of HII, HeII and HeIII
respectively. The ionization fractions are defined as: $x_{\rm
HII}=n_{\rm HII}/n_{\rm HI}$, $x_{\rm HeII}=n_{\rm HeII}/n_{\rm HeI}$
and $x_{\rm HeIII}=n_{\rm HeIII}/n_{\rm HeI}$.

The mean hydrogen and helium densities vary with redshift as
$\bar{n}_{\rm HI, HeI}=\bar{n}_{\rm HI(0), HeI(0)}(1+z)^3$, where
$\bar{n}_{\rm HI(0), HeI(0)}$ are the mean hydrogen and helium
densities at the present time: $\bar{n}_{\rm
HI(0)}=1.9\times10^{-7}~{\rm cm^{-3}}$ and $\bar{n}_{\rm
HeI(0)}=1.5\times10^{-8}~{\rm cm^{-3}}$.

By inserting Eq.~\ref{eq:ne} into Eq.~\ref{eq:dTkCMB} and converting
Eq.~\ref{eq:dTkCMB} from an integral in time to one in redshift space
\footnote{In order to make transformation of the Eq.~\ref{eq:dTkCMB}
to the redshift space we use ${\rm d}t=-\frac{{\rm d}z}{H(z)[1+z]}$,
where $H(z)$ is Hubble constant at redshift $z$.} ($z$), we get:

\begin{eqnarray}\label{eq:dTkCMBz}
\left( \frac{\delta T}{T} \right)_{\rm kSZ} &=& -{\rm \sigma_{T}}{\rm
  \bar{n}_{\rm n(0)}} \int_{z_{\rm r}}^{z_{0}} \frac{(1+z)^2}{H}{\rm
  e}^{-\tau} \bar{x_e} \nonumber \\ & &\cdot(1 +\delta + \delta_{x_e}
  + \delta \delta_{x_e} )v_{\rm r}{\rm d}z,
\end{eqnarray}
where $v_{\rm r}$ is component of $\vec{v}$ along the LOS
($v_r=\hat{r} \cdot \vec{v}$) and $\bar{n}_{\rm n(0)}=\bar{n}_{\rm
HI(0)}+\bar{n}_{\rm HeI(0)}$. For a $\Lambda$CDM Universe the Hubble
constant at redshift $z$ is $H={\rm H_0} \sqrt{{\rm
\Omega_m}(1+z)^3 + {\rm \Omega_\Lambda}}$ where ${\rm H_0}$ is the
present value of the Hubble constant, ${\rm \Omega_m}$ is the matter
and ${\rm \Omega_\Lambda}$ the dark energy densities, respectively.

For homogeneous reionization histories, i.e. a uniform change in the
ionization fraction as a function of redshift, Eq.~\ref{eq:dTkCMBz}
becomes:
\begin{equation}
  \left( \frac{\delta T}{T} \right)_{\rm kSZ} = -{\rm \sigma_{T}}{\rm
    \bar{n}_{\rm H(0)}} \int_{z_{\rm r}}^{z_{0}} \frac{(1+z)^2}{H}{\rm
    e}^{-\tau} \bar{x_e} (1 +\delta)v_{\rm r}{\rm d}z,
\end{equation}
which means that the kSZ fluctuations are induced only by spatial
variations of the density field. The linear regime of this effect is
called the Ostriker-Vishniac (OV) effect. The OV effect is of second
order and peaks at small angular scales (arc minutes) and
has an $rms$ of the order of a few ${\rm \mu K}$.

\subsection{The Cosmological 21~cm signal}
\label{sec:eor}

In radio astronomy, where the Rayleigh-Jeans law is applicable, the
radiation intensity, $I(\nu)$ is expressed in terms of the brightness
temperature $T_{b}$:
\begin{equation}
I(\nu) = \frac{2 \nu^2}{c^2} k T_b,
\end{equation}
where $\nu$ is the frequency, $c$ is the speed of light and $k$ is
Boltzmann's constant. The predicted differential brightness
temperature of the cosmological 21~cm signal with the CMB as the
background is given by \citep{field58, field59, ciardi03}:
\begin{eqnarray}
\delta T_{\rm b}&=&26~\mathrm{mK}~x_{\rm HI}(1+\delta)
 \left(1-\frac{T_{\rm CMB}}{T_{\rm s}}\right) \left(\frac{\Omega_{\rm
 b} h^2}{0.02}\right)\nonumber\\ &&\left[\left(\frac{1 +
 z}{10}\right)\left(\frac{0.3}{\Omega_{\rm m}}\right)\right]^{1/2}.
\label{eq:tbright0}
\end{eqnarray}
Here $T_{\rm s}$ is the spin temperature, $x_{\rm HI}$ is the neutral
hydrogen fraction, $\delta$ is the matter density contrast and
$h=H_{0}/(100~{\rm kms^{-1}Mpc^{-1}})$. If we express the neutral
hydrogen fraction as $x_{\rm HI}=\bar{x}_{\rm HI}(1+\delta_{x_{\rm
HI}})$, Eq.~\ref{eq:tbright0} becomes:
\begin{eqnarray}
\delta T_{\rm b} & = &26~\mathrm{mK}~\bar{x}_{\rm
 HI}(1+\delta+\delta_{x_{\rm HI}}+\delta\delta_{x_{\rm HI}})
  \nonumber\\ & &\left(1-\frac{T_{\rm CMB}}{T_{\rm s}}\right)
 \left(\frac{\Omega_{\rm b} h^2}{0.02}\right)\left[\left(\frac{1 +
 z}{10}\right)\left(\frac{0.3}{\Omega_{\rm m}}\right)\right]^{1/2}.
\label{eq:tbright}
\end{eqnarray}

In his two seminal papers, \citet{field58, field59} calculated the
spin temperature, $T_{s},$ as a weighted average of the CMB, kinetic
and colour temperatures:
\begin{equation}
T_{s} = \frac{T_{CMB} +y_{kin} T_{kin} + y_\alpha
T_{\alpha}}{1+y_{kin}+y_\alpha},
\label{eq:tspin}
\end{equation}
where $T_{CMB}$ is the CMB temperature and $y_{kin}$ and $y_\alpha$
are the kinetic and Lyman-$\alpha$ coupling terms, respectively. We
have assumed that the color temperature, $T_{\alpha}$, is equal to
$T_{kin}$ \citep{madau97}. The kinetic coupling term increases with
the kinetic temperature, whereas the $y_\alpha$ coupling term depends
on Lyman-$\alpha$ pumping through the so-called Wouthuysen-Field
effect \citep{wouthuysen52, field58}. The two coupling terms are
dominant under different conditions and in principle could be used to
distinguish between ionization sources, e.g., between first stars, for
which Lyman-$\alpha$ pumping is dominant, vs. first mini-quasars for
which X-ray photons and therefore heating is dominant \citep[see
e.g.,~][]{madau97, zaroubi07, thomas08}.

\section{Simulations}\label{sec:sim}
The kSZ ($\delta T/T$) and the cosmological 21~cm maps ($\delta T_b$)
are simulated using the following data cubes: density ($\delta$),
radial velocity ($v_r$) and HI, HII, HeI, HeII and HeIII fractions
($x_{\rm HI, HII, HeI, HeII \& HeIII}$). The data cubes are
produced using the \textsc{bears} algorithm, a fast algorithm to
simulate the EoR signal \citep{thomas09}.

In the following subsections we summarize the \textsc{bears} algorithm
and describe operations done on the output in order to calculate the
kSZ and the EoR maps. Furthermore, we detail the calculations to
obtain the optical depth and kSZ signal along a certain LOS. Finally
we present the maps of the kSZ temperature fluctuations for the two
patchy reionization models (`stars' and `mini-quasars') and discuss
aspects of their contribution to the signal.

\subsection{BEARS algorithm: overview}
\textsc{bears} is a fast algorithm to simulate the underlying
cosmological 21~cm signal from the EoR. It is implemented by using an
N-body/SPH simulation in conjunction with a 1-D radiative transfer
code under the assumption of spherical symmetry of the ionized
bubbles. The basic steps of the algorithm are as follows: first, a
catalogue of 1D ionization profiles of all atomic hydrogen and helium
species and the temperature profile that surround the source is
calculated for different types of ionizing sources with varying
masses, luminosities at different redshifts. Subsequently, photon
rates emanating from dark matter haloes, identified in the N-body
simulation, are calculated semi--analytically. Finally, given the
spectrum, luminosity and the density around the source, a spherical
ionization bubble is embedded around the source, whose radial profile
is selected from the catalogue as generated above. For more details
refer \citet{thomas09}.

As outputs we obtain data cubes (2D slices along the
frequency/redshift direction) of density ($\delta$), radial velocity
($v_r$) and hydrogen and helium fractions ($x_{\rm HI, HII, HeI, HeII
\& HeIII}$).  Each data cube consists of about 850 slices each
representing a certain redshift between 6 and 11.5. Slices have a size
of $100~\rm{h^{-1}}$ comoving Mpc and are defined on a $512^2$
grid. Becauses these slices are produced to simulate a mock dataset
for radio-interferometric experiments, they are uniformly spaced in
frequency (therefore, not uniform in redshift). Thus, the frequency
resolution of the instrument dictates the scales over which structures
in the Universe are averaged/smoothed along the redshift
direction. The relation between frequency $\nu$ and redshift space $z$
is given by:
\begin{equation}
  z=\frac{\nu_{21}}{\nu}-1,
\end{equation}
where $\nu_{21}=1420~\rm{MHz}$ is the rest frequency that corresponds
to the 21cm line.

The final data cubes are produced using approximately 35 snapshots of
the cosmological simulations. Since choice of the redshift direction
in each box is arbitrary, three final data cubes can be produced in
this manner (x, y and z).

\subsection{Randomization of the structures}
The kSZ effect is an integrated effect and is sensitive to the
structure distribution along the LOS.  To avoid unnatural
amplification of the kSZ fluctuations due to repeating structures in
the simulated data cubes, we follow \citet{iliev07} and introduce
randomization of the structures along the LOS over $100~{\rm Mpc/h}$
scale in two steps. First, each $100~{\rm Mpc/h}$ chunk of the data
cube is randomly shifted (assuming periodic boundary conditions) and
rotated in a direction perpendicular to the LOS. The shift can be
positive or negative in any direction ($x$ and(or) $y$) by an integer
value between 0 and 512. While the rotation can be clockwise or
anticlockwise by an $n\pi/2$ angle (n=0,1,2,3). Secondly the final
data cube is produced by assembling the first $100~{\rm Mpc/h}$ part
from the x-data cube, second from the y-data cube, third from the
z-data cube and then back to the x-data cube and so on., to a distance
that spans the comoving radial distance between redshifts 6 and 12.

\subsection{Optical depth}
Thomson optical depth $\tau$ at redshift $z$ is:
\begin{equation}\label{eq:tau}
  \tau={\rm c \sigma_T} \int_{0}^{z} n_{\rm e}\frac{(1+z)^2}{H(z)}{\rm
  d}z,
\end{equation}
where ${\rm c}=2.998 \times 10^{8}~{\rm m~s^{-1}}$ is the speed of
light, ${\rm \sigma_T}=6.65\times10^{-29}~{\rm m^2}$ the Thomson
scattering cross section for electrons, $n_{\rm e}$ the density of
free electrons and $H(z)$ the Hubble constant at redshift $z$.

In our simulations we split the integral into two parts. The first
part that represents the mean Thomson optical depth
($\bar{\tau}_{06}$) between the redshift 0--6 and the second,
$\tau_{6z}$, from redshift 6 to a desired redshift $z$. This choice is
driven by the limited redshift range ($z\sim6-11.5$) of imminent radio
astronomical projects designed to map the EoR. Under the assumption
that the reionization is completed by the redshift 6, the mean Thomson
optical depth $\bar{\tau}_{06}$ is $0.0517$. Note that our simulation
is set to have a mean Thomson optical depth of 0.087, as obtained from
the CMB data ($\tau=0.087\pm0.017$, \citet{komatsu08}).

\subsection{Creating the kSZ and EoR maps}

\begin{figure*}
\centering \includegraphics[width=.9\textwidth]{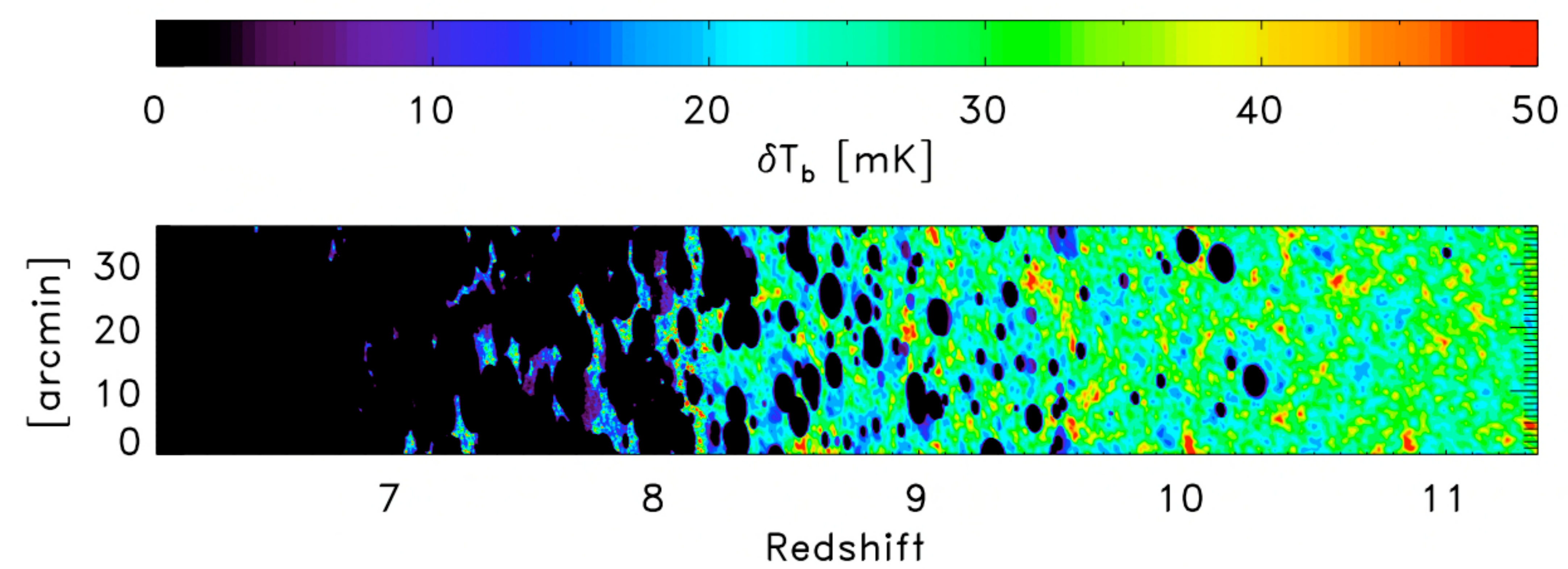}
\centering \includegraphics[width=.9\textwidth]{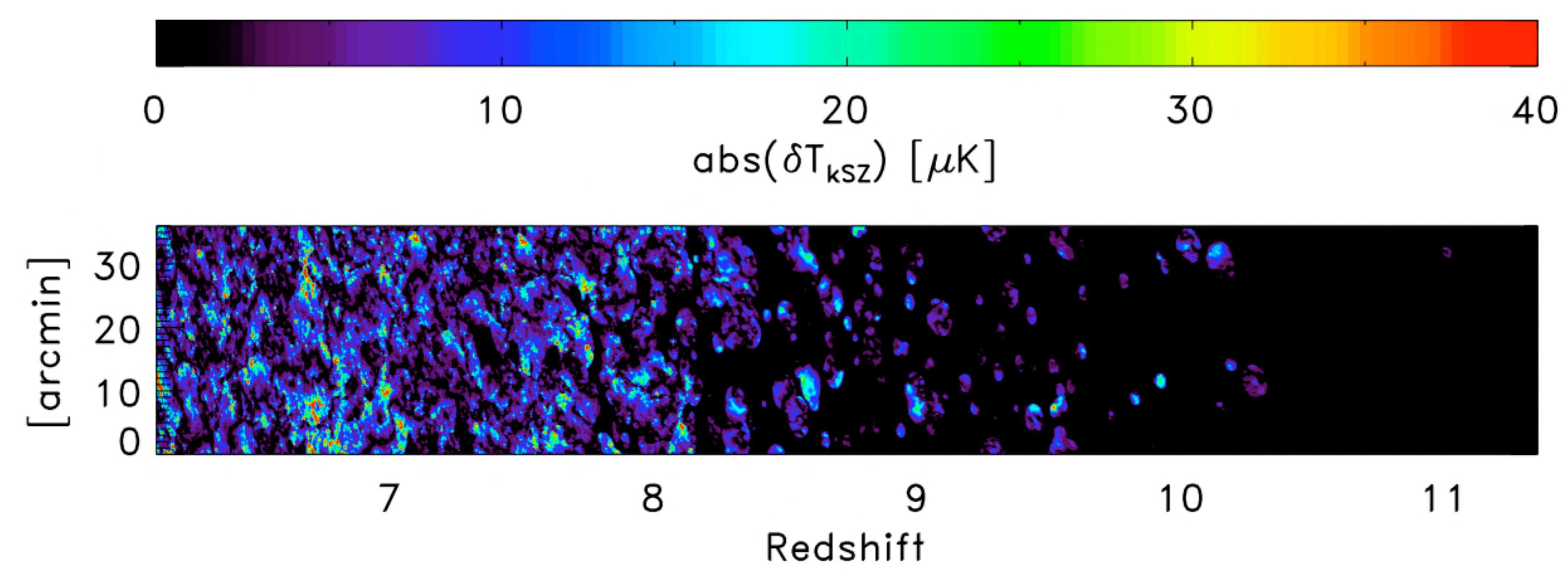}
\caption{\emph{The same as a Fig.~\ref{fig:EoRkSZq} but for the `QSOs'
patchy reionization model.}}
\label{fig:EoRkSZq}
\end{figure*}

For clarity we summarize the steps we follow to create the kSZ and EoR
maps for a given scenario of the reionization history:
\begin{enumerate}
\item Using the output of \textsc{bears}, data cubes for the density,
      radial velocity, helium and hydrogen fractions are produced.
\item Data cubes are randomized over $100~{\rm Mpc/h}$ scale along
      the redshift direction.
\item Using Eq.~\ref{eq:tau} the Thomson optical depth, $\tau$, is 
      calculated to a redshift $z$.
\item Using the integrand of the Eq.~\ref{eq:dTkCMBz}, data cubes 
      with the kSZ signal is produced as a function of redshift.
\item Integrating along each LOS through the kSZ data cube
      the integrated kSZ map is obtained. Note that we assume that the
      reionization is complete by redshift 6, so the integral in
      Eq.~\ref{eq:dTkCMBz} spans $z>6$.
\item Finally, the brightness temperature fluctuations, $\delta T_b$,
      is calculated using Eq.~\ref{eq:tbright}.
\end{enumerate}

In the following sections we will use the kSZ and EoR maps produced
for three different models of reionization:
\begin{enumerate}
  \item \textsc{homogeneous:} Reionization history is homogeneous and
  ionized fraction follows:
  \begin{equation}\label{eq:hhist}
    x_e=\frac{1}{1+e^{k(z-z_{\rm reion})}},
  \end{equation}
  with $z_{\rm reion}$ set to 8.5 and $k=1,2,...$ tunes the
  ``rapidness'' of the reionization process.
\item \textsc{patchy Stars:} Reionization history is patchy, gradual
  and extended with stars as the sources of ionization.
  \item \textsc{patchy QSOs:} Reionization history is patchy and
  relatively fast with QSOs as the ionizing sources.
\end{enumerate}

Fig.~\ref{fig:EoRkSZs} \& Fig.~\ref{fig:EoRkSZq} show slices through
the simulated redshift cube of the cosmological 21~cm signal ($\delta
T_{\rm b}$) and the kSZ effect ($\delta T_{\rm kSZ}$) in the case of
`Stars' and `QSOs' patchy reionization models. The angular size of the
slices is $\sim 0.6^\circ$.

\begin{figure}
\centering \includegraphics[width=.45\textwidth]{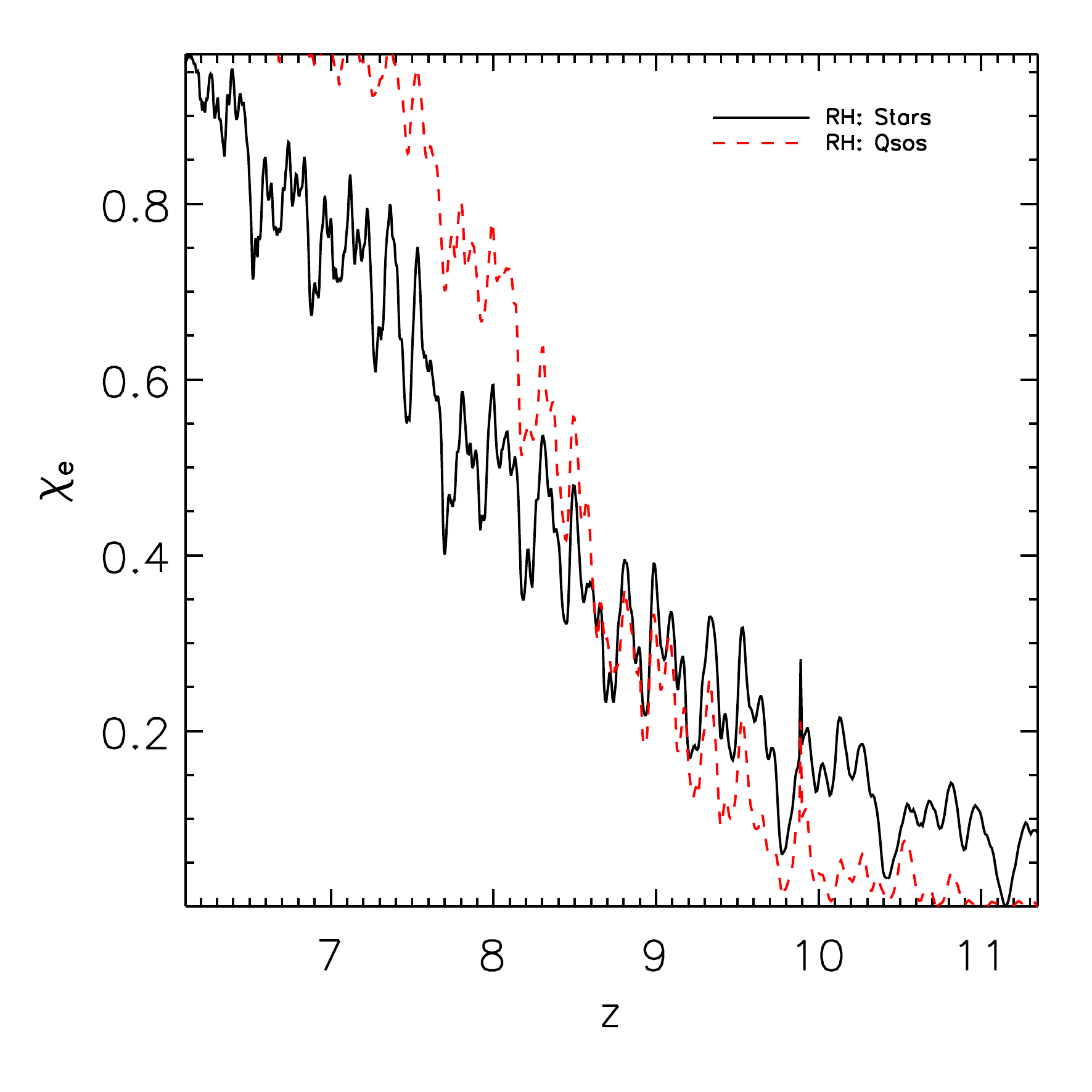}
\caption{\emph{The mean ionization fraction $x_{\rm e}$ as a
    function of redshift for the `Stars' and `QSOs' patchy
    reionization model.}}
\label{fig:RHsq}
\end{figure}

Apart from the difference in the global shape of the reionization
histories driven by `Stars' and `QSOs' (see Fig.~\ref{fig:RHsq}), the
average size of the ionization bubble is also smaller in `Stars'
compared to that of `QSOs'. For a detailed description and comparison
of reionization histories due to `Stars' and `QSOs' see
\citet{thomas09}.

\begin{figure*}
\centering \includegraphics[width=.24\textwidth]{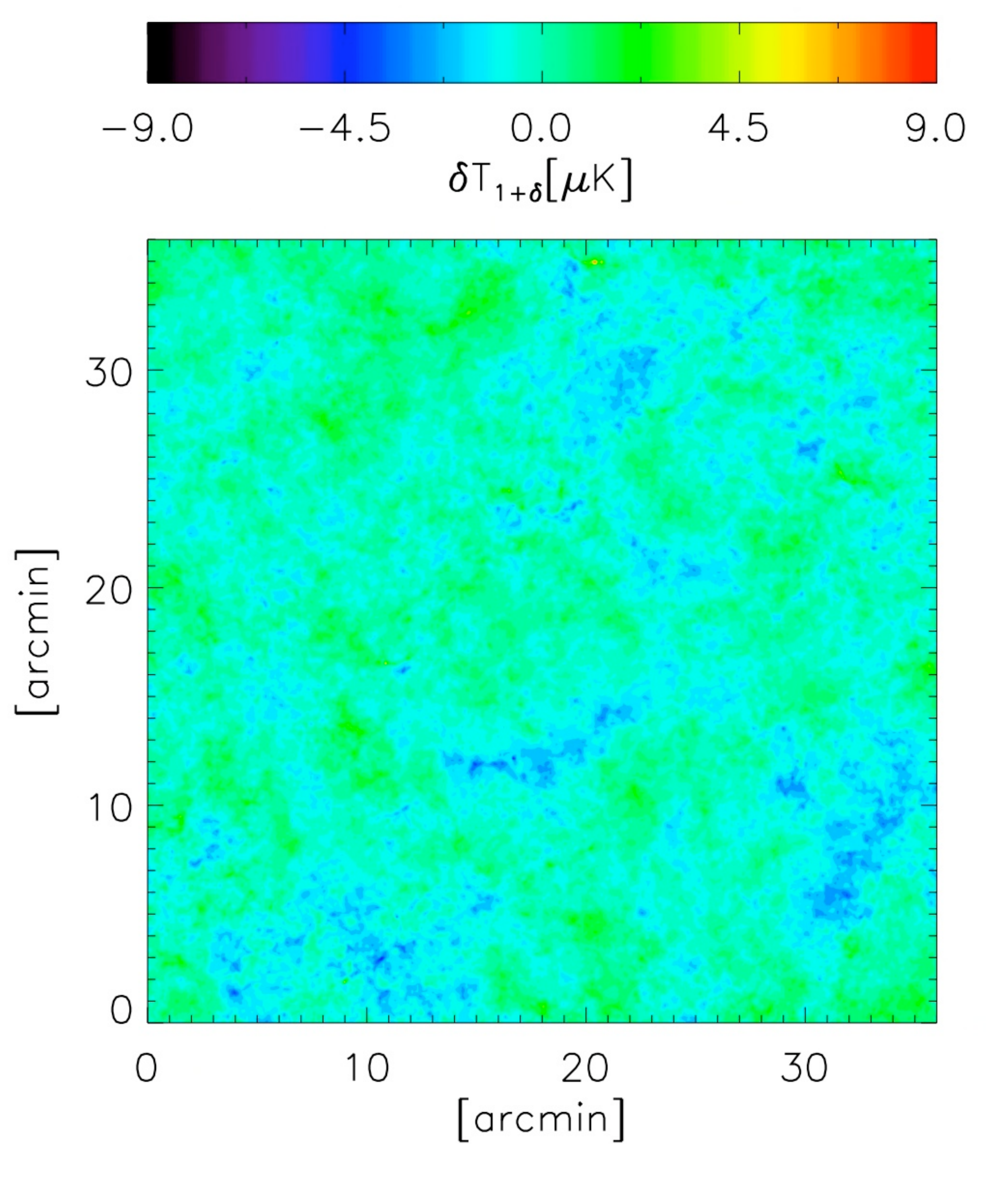}
\centering \includegraphics[width=.24\textwidth]{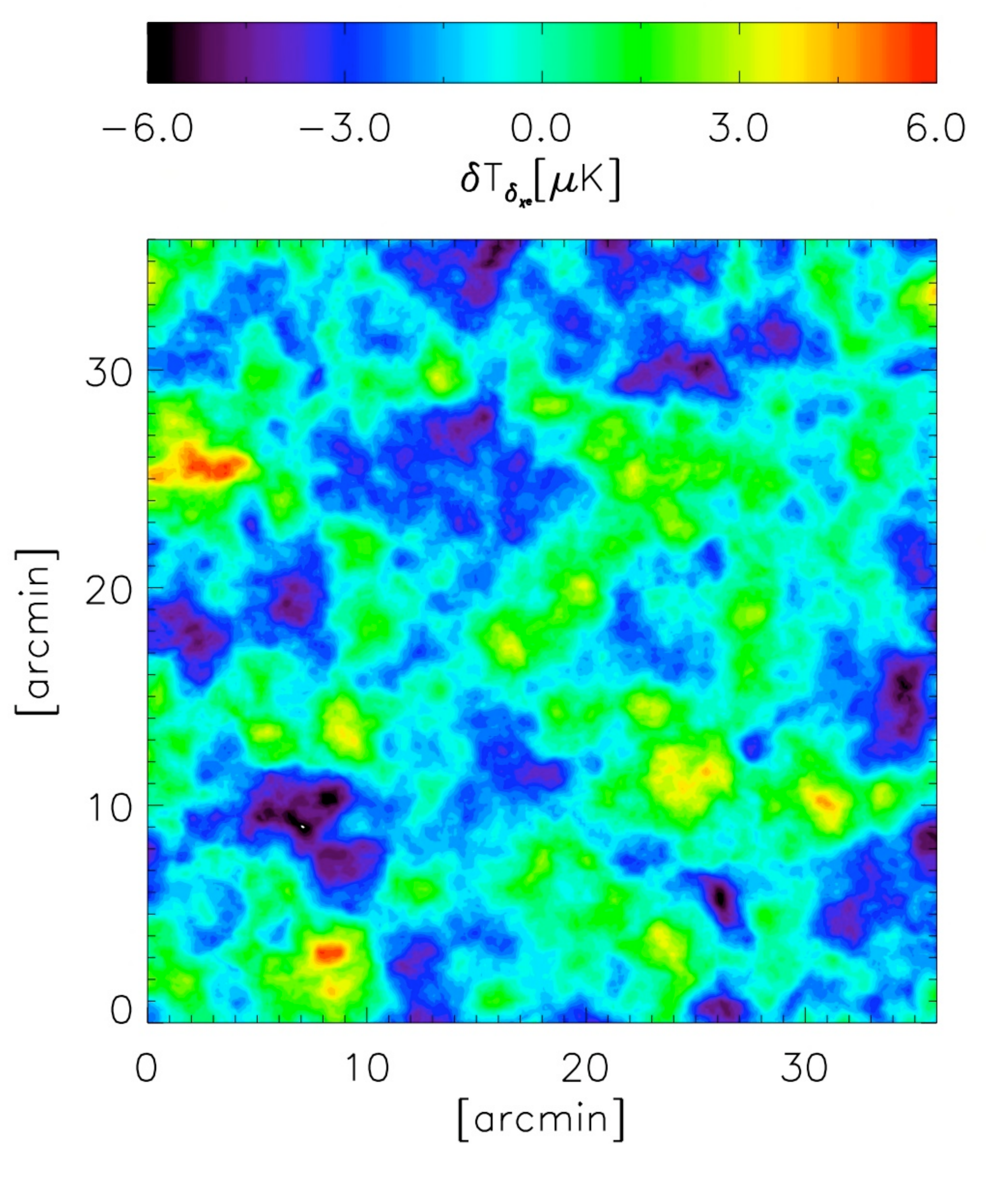}
\centering \includegraphics[width=.24\textwidth]{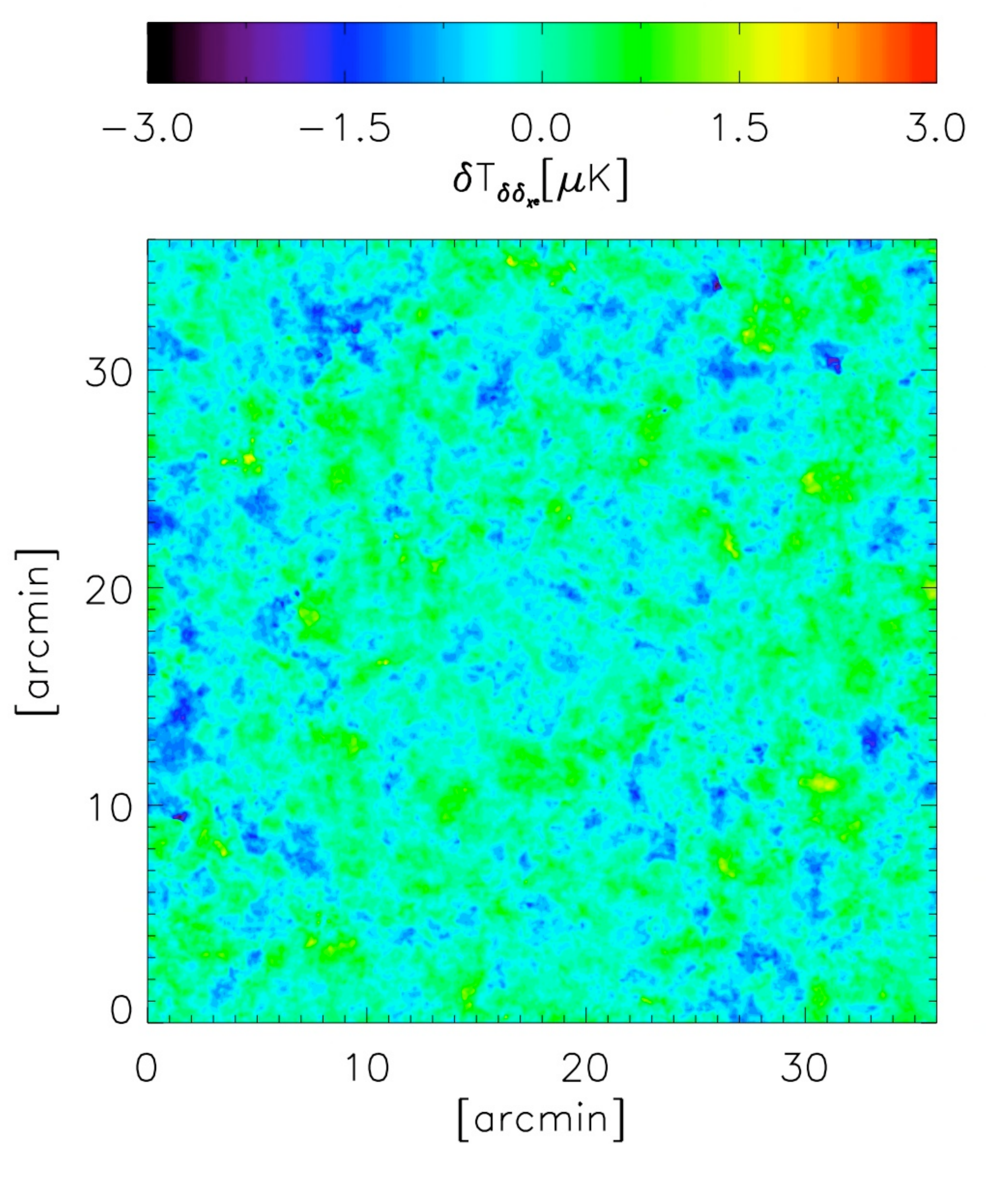}
\centering \includegraphics[width=.24\textwidth]{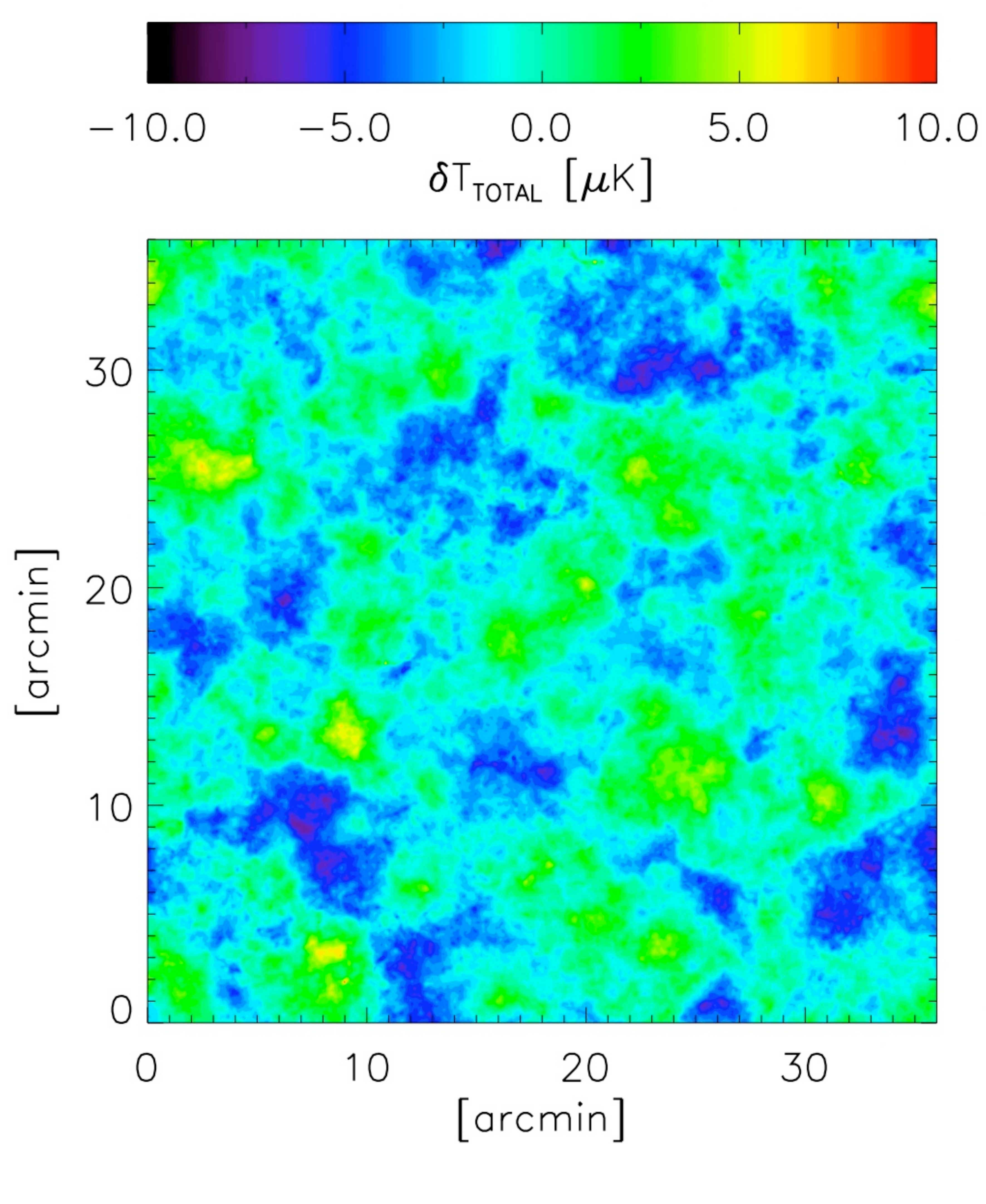}
\caption{\emph{The simulated kSZ anisotropies induced by `$1+\delta$'
    (first panel), `$\delta_{x_{\rm e}}$' (second panel) and
    `$\delta\delta_{x_{\rm e}}$' term (third panel) in
    Eq.~\ref{eq:dTkCMBz} for patchy reionization model by `Stars'. The
    kSZ anisotropies induced by all terms together in
    Eq.~\ref{eq:dTkCMBz} are showed on fourth panel
    (`\textsc{total}'). The ${\rm mean}$ and ${\rm rms}$ of the
    simulated kSZ maps is shown in Table~\ref{tab:rmsSQ}. Note that
    each map has its own color scale.}}
\label{fig:kSZcontS}
\end{figure*}

The kSZ anisotropies from patchy reionization is induced by both
fluctuations of the density field $\delta$ and ionization fraction
$\delta_{x_{\rm e}}$ (see Eq.~\ref{eq:dTkCMBz}). \citet{santos03}
found that kSZ anisotropies from $\delta_{x_{\rm e}}$ fluctuations
dominate over $\delta$ modulated fluctuations (OV effect). In order to
test this result with our simulations we split the integral in
Eq.~\ref{eq:dTkCMBz} into three parts and produce three integrated kSZ
maps (for the `Stars' model see Fig.~\ref{fig:kSZcontS}). The first
term `$1+\delta$' represents the density induced secondary
anisotropies (OV effect). The `$\delta_{x_{\rm e}}$' term represents
the secondary anisotropies due to patchiness in the reionization and
`$\delta\delta_{x_{\rm e}}$' represents a higher order anisotropy.

\begin{table}
  \centering
  \caption{The $mean$ and $rms$ of the `$1+\delta$', `$\delta_{x_{\rm
    e}}$' and `$\delta\delta_{x_{\rm e}}$' simulated kSZ maps for both
    `Stars' (see Fig.~\ref{fig:kSZcontS}) and `QSOs' patchy
    reionization model. $C_0$ is a cross-correlation coefficent at a
    zero lag between coresponding kSZ maps and integrated EoR map (see
    Section~\ref{sec:cross}). For completeness we also show results
    for the pure Doppler term (`$1$') in Eq.~\ref{eq:dTkCMBz}.}
  \begin{tabular}{@{}ccccccc@{}}
    \hline & $\delta T_{\rm kSZ}$ & `$1$' & `$1+\delta$' &
    `$\delta_{x_{\rm e}}$' & `$\delta\delta_{x_{\rm e}}$' &
    \textsc{total} \\ \hline\textsc{stars} & $mean$ [${\rm \mu K}$]
    &-0.004& 0.03 & 0.58 & 0.02 & 0.63 \\ & $rms$ [${\rm \mu K}$] &
    0.14 & 0.80 & 1.74 & 0.40 & 2.00 \\ & $C_{0}$ & 0.05 & -0.003 &
    -0.12 & -0.06 & -0.11\\\hline \textsc{qsos} & $mean$ [${\rm \mu
    K}$] & -0.002 & 0.03 & 0.27 & 0.01 & 0.30 \\ & $rms$ [${\rm \mu
    K}$]& 0.15 & 0.93 & 1.28 & 0.28 & 1.57 \\ & $C_{0}$ & 0.1 & 0.04 &
    -0.08 & -0.01 & -0.06 \\\hline
  \end{tabular}\label{tab:rmsSQ}
\end{table}

The $mean$ and $rms$ of the `$1+\delta$', `$\delta_{x_{\rm e}}$' and
`$\delta\delta_{x_{\rm e}}$' components of the simulated kSZ maps are
shown in the Table~\ref{tab:rmsSQ} for patchy reionization in the
`Stars' and `QSOs' model. The $rms$ value of the maps is used as
a measure of the fluctuations. We confirm that the `$\delta_{x_{\rm e}}$'
fluctuations are indeed larger than density induced anisotropies (`$\delta$')
for both patchy reionization models. However the difference between
the `$\delta_{x_{\rm e}}$' and `$\delta$' fluctuations is much larger
for the `Stars'reionization history model than `QSOs' model. Also note
that the third order anisotropy (`$\delta\delta_{x_{\rm e}}$') is not
negligible in both reionization scenarios. For completeness we also
show contribution from the pure Doppler term (`$1$') in
Eq.~\ref{eq:dTkCMBz}.

\section{Cross-correlation kSZ-EoR maps}\label{sec:cross}
The kSZ effect from the EoR is expected to be correlated with
cosmological 21~cm maps for a homogeneous reionization history and
anti-correlated when patchy \citep{cooray04, salvaterra05, alvarez06,
slosar07, adshead08}. In this section simulations described in
Section~\ref{sec:sim} are used to explore the small angular scale
cross-correlation between the kSZ effect and EoR maps for five
different reionization histories. Further, we will fold-in the
influence of i) the large-scale velocities on the kSZ effect and ii)
the primary CMB fluctuations on the cross-correlation.

Throughout the paper we will use a normalized cross-correlation in
order to be able to compare results from a different pairs of
maps. The normalized cross-correlation between two images ($a_{i,j}$
and $b_{i,j}$) with the same total number of pixels $n$ is defined at
zero lag as:
\begin{equation}\label{eq:c0}
C_0=\frac{1}{n-1}\sum_{i,j} \frac{(a_{i,j} - \bar{a})
(b_{i,j}-\bar{b})}{\sigma_{a}\sigma_{b}},
\end{equation}
where $\bar{a}$ ($\bar{b}$) is the mean and $\sigma_{a}$
($\sigma_{b}$) the standard deviation of the image $a$ ($b$). However,
the cross-correlation between the kSZ and the EoR map needs to be
considered more carefully.

The fluctuations of the kSZ effect over the simulated map are both
positive and negative, since the radial velocity $v_{\rm r}$ can be
both positive and negative (see Eq.~\ref{eq:dTkCMBz}). In contrast,
the EoR signal fluctuations in our simulations are always positive
(see Eq.~\ref{eq:tbright}). When calculating the cross-correlation
between these two maps, we are interested in finding the number of
points at which both signals are present (homogeneous reionization
model) or where one signal is present and the other absent (patchy
reionization model). In other words only the absolute value of the kSZ
fluctuation is relevant in our calculation and not its sign.

\subsection{Homogeneous reionization history}
\begin{figure}
\centering \includegraphics[width=.45\textwidth]{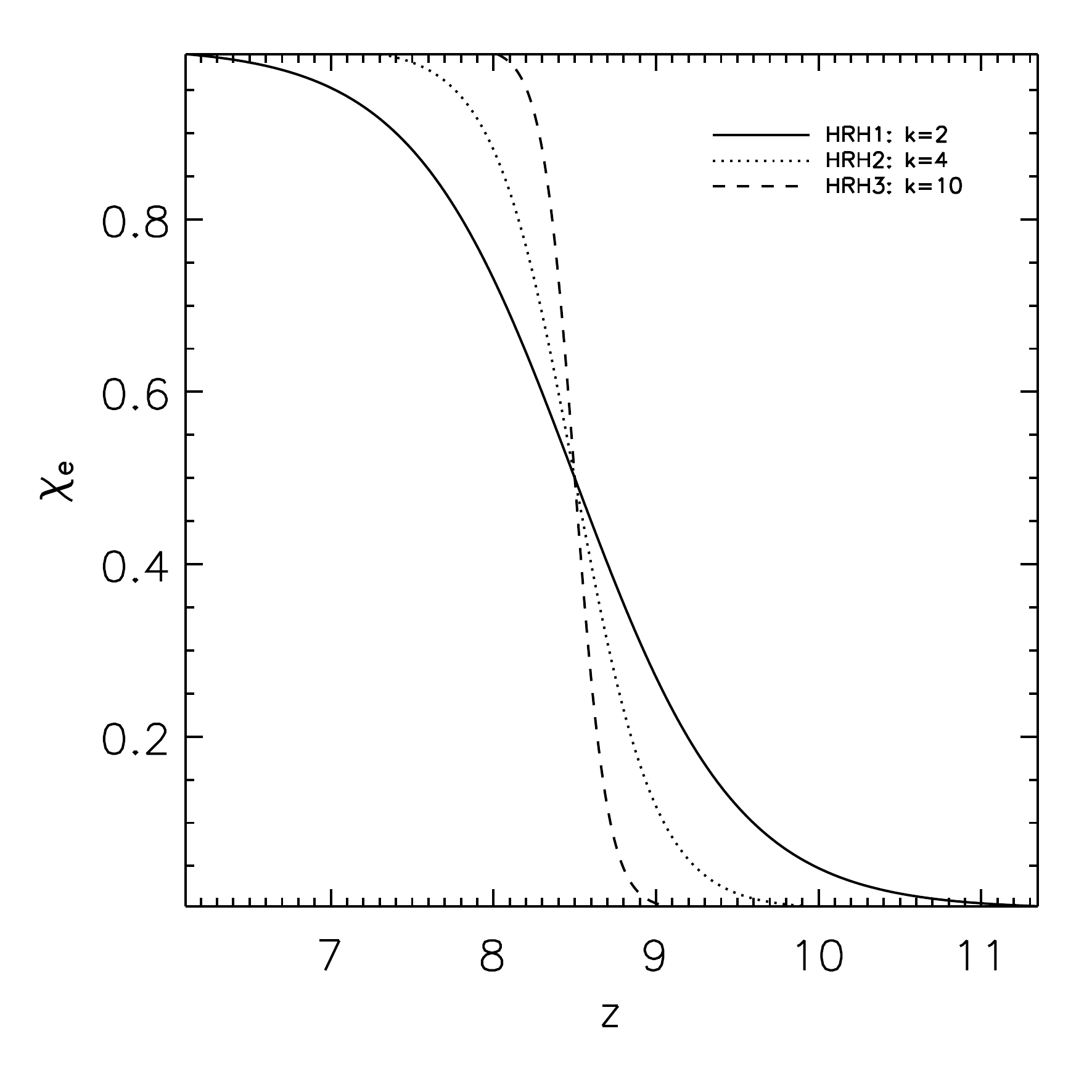}
\caption{\emph{The ionization fraction $x_{\rm e}$ as a function of
    redshift for three different models of the homogeneous
    reionization history (\textsc{hrh 1}, \textsc{hrh 2} \&
    \textsc{hrh 3}). All three models are defined by
    Eq.~\ref{eq:hhist} but have different values of $k$ (different
    reionization durations). The half of the reionization process is
    set in all three models to $z_{reion}=8.5$.}}
\label{fig:HRH}
\end{figure}

We explore the cross-correlation between the kSZ map and integrated
EoR map in the case of three different homogeneous reionization
histories (HRH1, HRH2 \& HRH3). These histories are modulated by
Eq.~\ref{eq:hhist}, with $k$ controlling the duration of
reionization. The mean ionization fraction $x_{\rm e}(z)$ for these
models are shown in Fig.~\ref{fig:HRH}.

The cross correlation between an integrated kSZ map and an integrated
EoR map results in a coefficient $C_{0, \rm \textsc{hrh
1}}=0.10\pm0.03$ for an extended homogeneous reionization history
(\textsc{hrh 1}).  For \textsc{hrh 2} $C_{0, \rm\textsc{hrh
2}}=0.21\pm0.02$ and \textsc{hrh 3} $C_{0, \rm \textsc{hrh
3}}=0.24\pm0.02$. The errors are estimated by performing a Monte Carlo
calculation with 200 independent realizations of the integrated kSZ
and EoR maps using the randomization procedure explained in
Section~\ref{sec:sim}.

As expected, the integrated kSZ and EoR maps are correlated for
homogeneous models of reionization. Furthermore, the correlation
depends on the duration of reionization with larger values for more
`rapid' reionization. These results are in agreement with
\citet{alvarez06}.

\subsection{Patchy reionization history}
\begin{figure}
\centering \includegraphics[width=.45\textwidth]{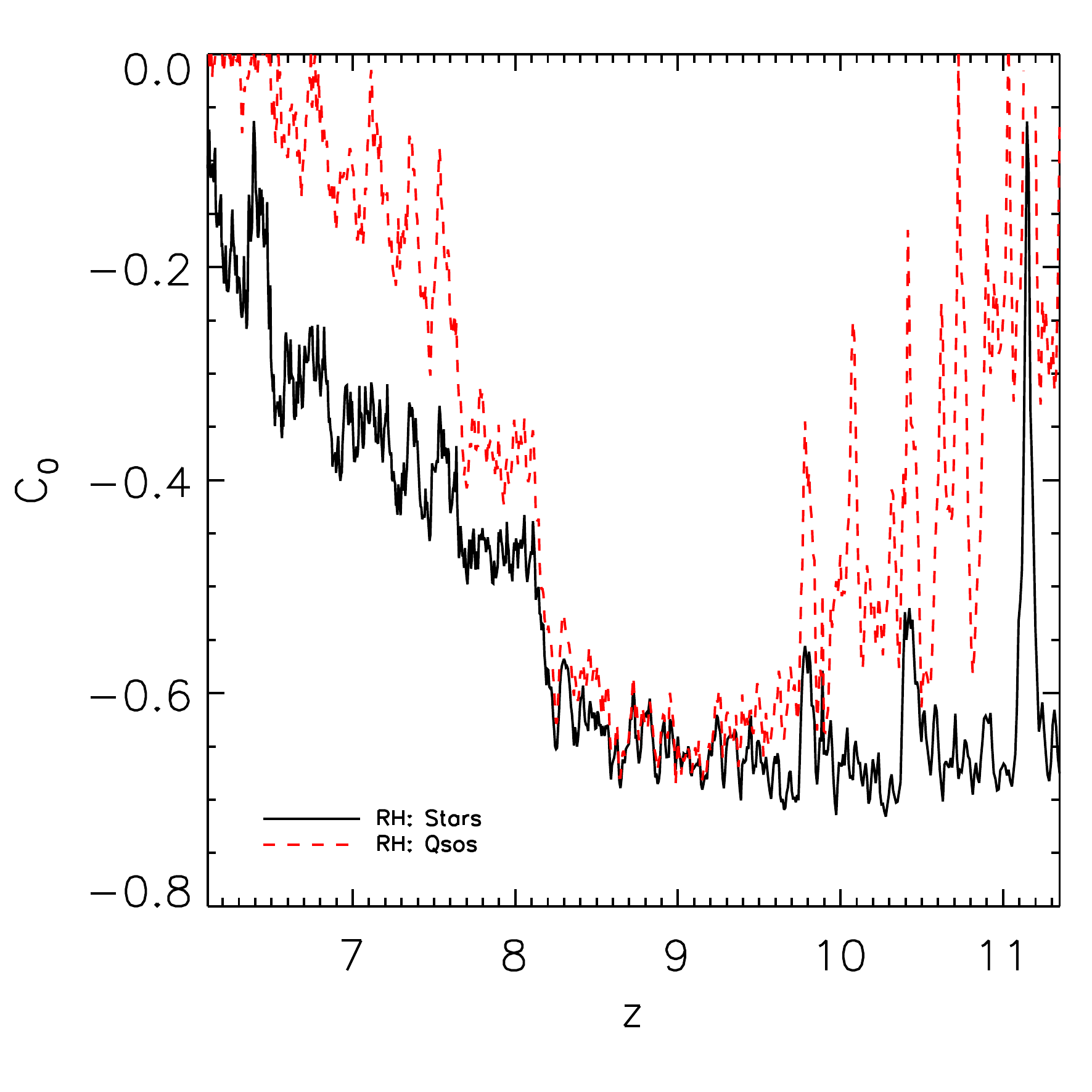}
\caption{\emph{The zero lag cross-correlation coefficient ($C_0$)
    between the kSZ map and the EoR map at a given redshift. The solid
    black line corresponds to the `Stars' while the dashed red line
    the `QSOs' patch reionization model. For both reionization models
    we find an anti-correlation between the maps.}}
\label{fig:corr0z}
\end{figure}

For patchy reionization models we first cross-correlate the kSZ and
the EoR map at a given redshift. The resulting zero lag coefficient
($C_0$), as a function of redshift, is shown in Fig.~\ref{fig:corr0z}.
The solid black line represents the correlation for `Stars' while the
dashed red line the `QSOs' patchy reionization model. As expected for
patchy reionization in both models, the kSZ and the EoR map
anti-correlate at individual redshifts.

Note that Eq.~\ref{eq:c0} is not taking into account a statistical
weight (significance) of the result. The statistical weight can be
defined with a factor $2n_*/n$ where $n_*$ represents the number of
$a_{i,j} \cdot b_{i,j}$ products that are non-zero. This makes the
result the most significant when both signals are represented equally
and less significant when one signal is present and the other not.

The obtained anti-correlation is also evident by visual inspection of
the kSZ and EoR slices through the simulated redshift cubes (see
Fig.~\ref{fig:EoRkSZs}~\&~\ref{fig:EoRkSZq}). One can see that the kSZ
signal is present only at the regions where the EoR signal is
not. This result is not surprising since the EoR signal is
proportional to neutral hydrogen while the kSZ to the ionized, both of
which are almost mutually exclusive.

In reality, we are not able to measure the kSZ effect at a certain
redshift but only an integrated effect along the entire history. Thus
we can only cross-correlate the integrated kSZ map with the integrated
EoR map and/or the EoR maps at different redshifts \footnote{This is
because, unlike the kSZ effect, we can potentially obtain
redshift-specific information of neutral hydrogen via upcoming radio
telescopes.}.

\begin{figure*}
\centering \includegraphics[width=.24\textwidth]{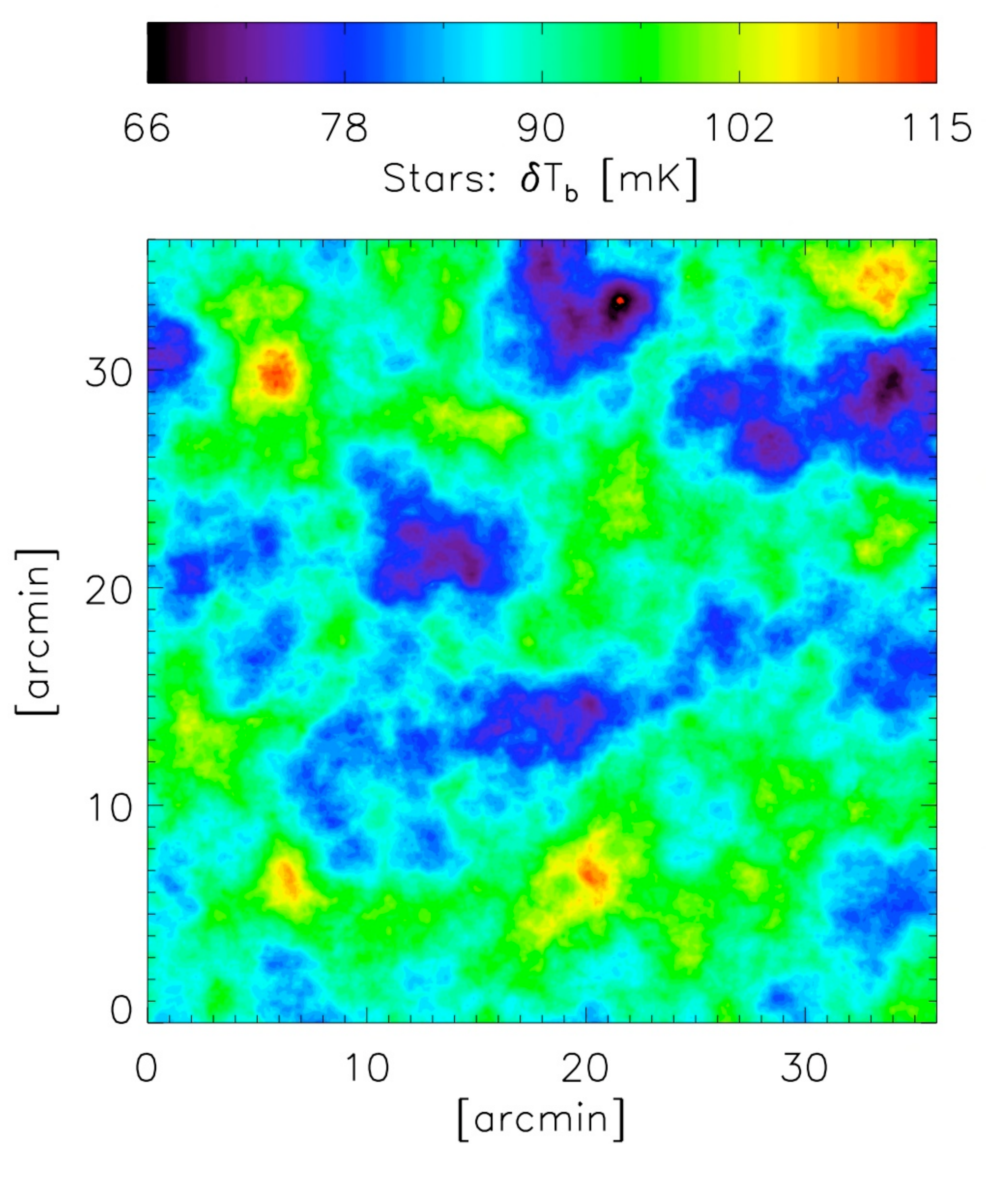}
\centering \includegraphics[width=.24\textwidth]{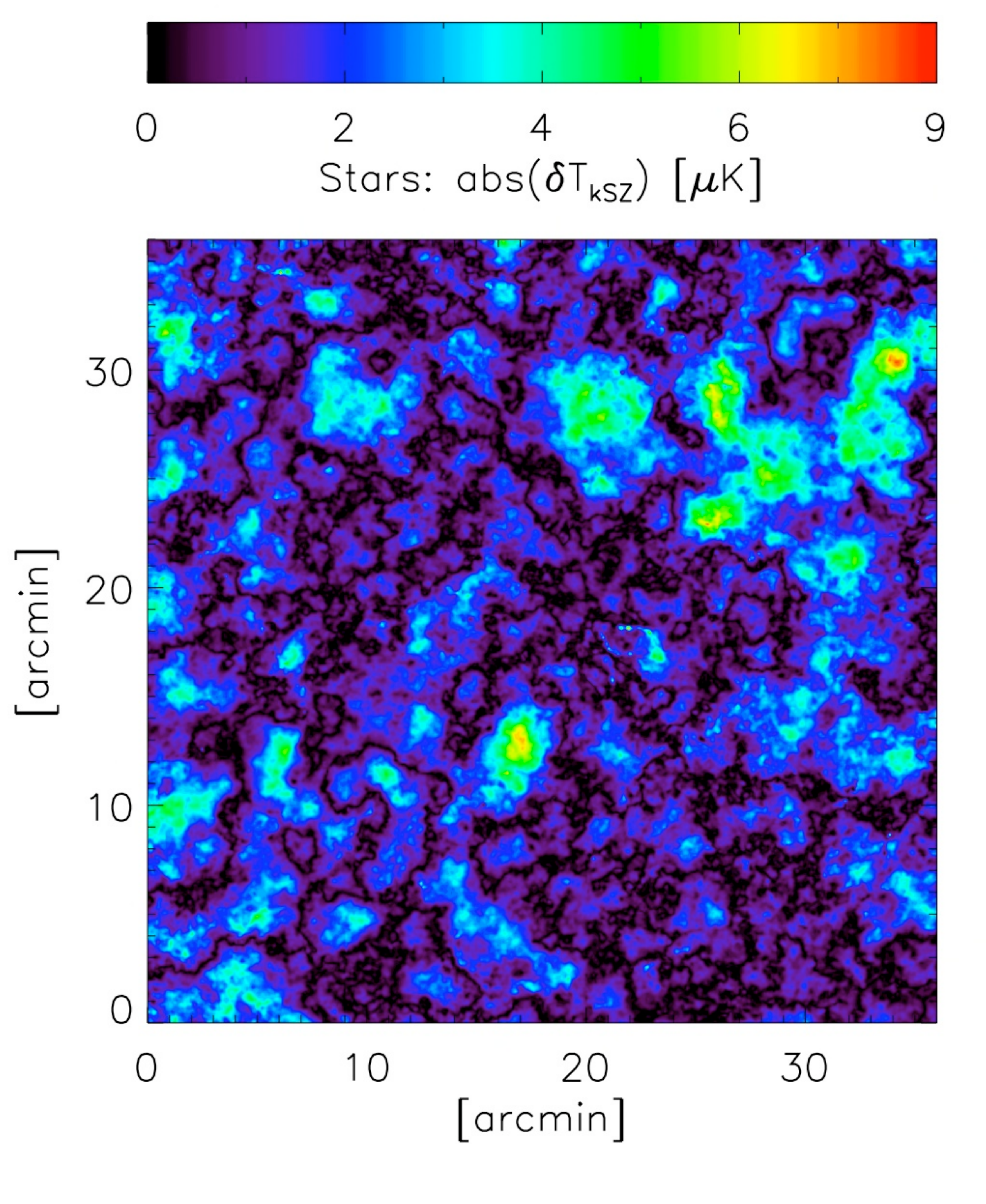}
\centering \includegraphics[width=.24\textwidth]{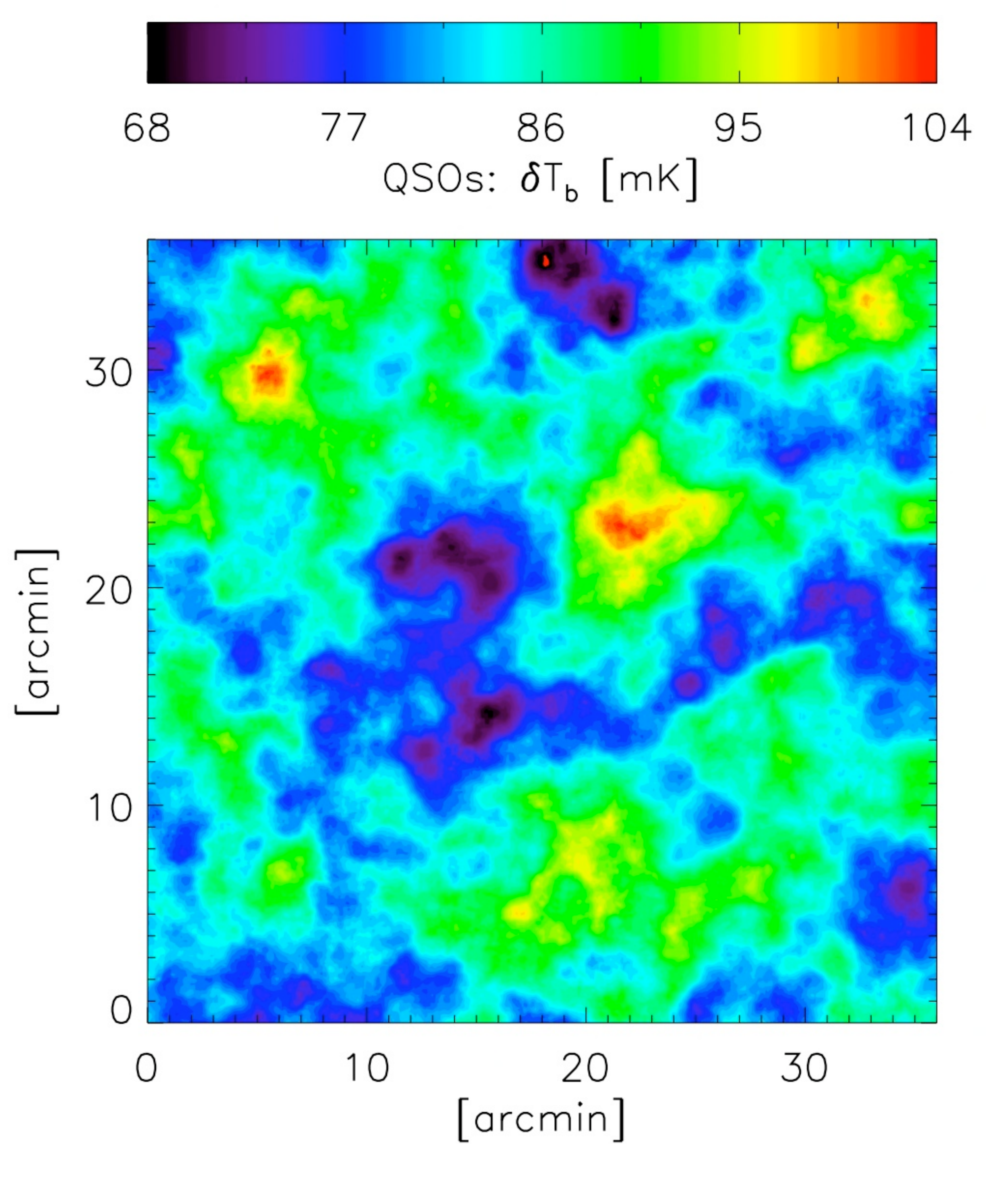}
\centering \includegraphics[width=.24\textwidth]{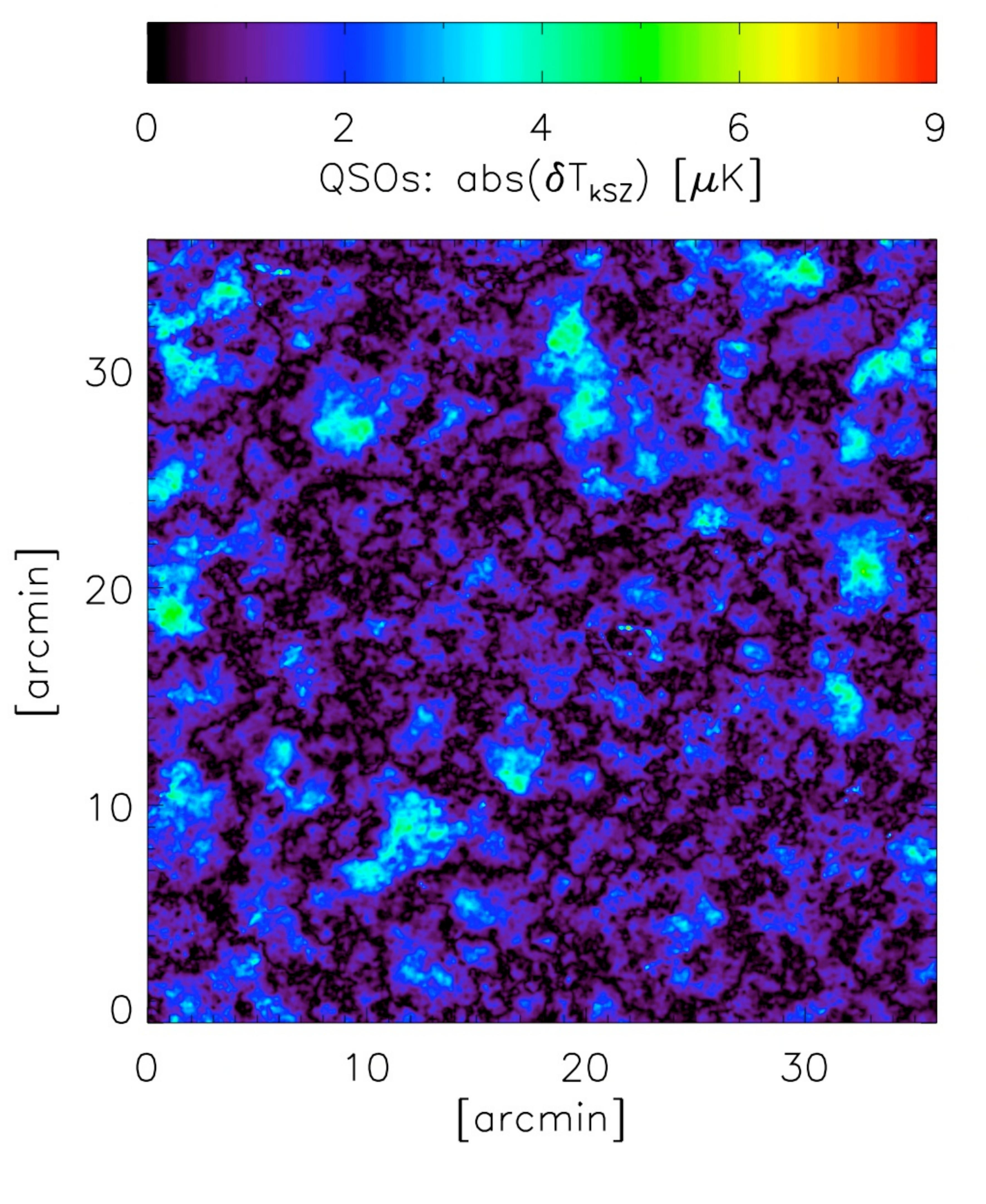}
\caption{\emph{The integrated EoR and kSZ map for the `Stars' (first
two panels) and `QSOs' patchy reionization model (second two
panels). The mean cross-correlation coefficient at the zero lag
between integrated EoR map and integrated kSZ map is $C_{0,{\rm
Stars}}=-0.16\pm0.02$ for the `Stars' and $C_{0,{\rm
QSOs}}=-0.05\pm0.02$ for the `QSOs' model.}}
\label{fig:EoRkSZinteg}
\end{figure*}

Fig.~\ref{fig:EoRkSZinteg} shows the integrated EoR and kSZ map for
the `Stars' (first two panels) and `QSOs' (last two panels) patchy
reionization model . The cross-correlation coefficient at zero lag for
these two models are $C_{0,{\rm Stars}}=-0.17$ and $C_{0,{\rm
QSOs}}=-0.02$. In order to determine the error on the kSZ-EoR
cross-correlation, we perform a Monte Carlo calculation. After
creating 200 independent realizations of the integrated kSZ and EoR
maps using the randomization procedure explained in
Section~\ref{sec:sim}, we calculate the cross-correlation coefficient
for each pair of realizations. Finally, we calculate the mean and
standard deviation of the cross-correlations. For the `Stars' model we
get $C_{0,{\rm Stars}}=-0.16\pm0.02$, while for the `QSOs' model
$C_{0,{\rm QSOs}}=-0.05\pm0.02$.

To understand higher values of the cross-correlation coefficient in
`Stars' compared to the `QSOs' model, one needs analyze
Fig.~\ref{fig:RHsq} and Table~\ref{tab:rmsSQ}. It is evident from
Fig.~\ref{fig:RHsq} that the reionization history is gradual and
extended with stars as ionizing sources, compared to a shorter and
sharper history with QSOs as ionizing sources. Moreover, the patchy
term (`$\delta_{x_{\rm e}}$') of the kSZ fluctuations is much larger
than the homogeneous component in the anisotropy (`$\delta$') in the
case of `Stars' than for `QSOs' model (see Table~\ref{tab:rmsSQ}).  We
showed beforehand that the kSZ effect correlates with the cosmological
21~cm signal for homogeneous reionization and that the correlation is
strongest for an `instant' reionization history. We also obtain the same result
by correlating different kSZ components with the
integrated EoR map (see Table~\ref{tab:rmsSQ}). Combining these
results we see the cross-correlation is driven by the patchy kSZ
anisotropies in the `Stars' model, while in the `QSOs' model the
homogeneous and patchy kSZ anisotropies tend to cancel each other. As
a consequence, the anti-correlation in 'QSOs' model is much weaker
than that of `Stars'.

In addition to the balance between homogeneous and patchy kSZ
anisotropies that governs the (anti-)correlation between the kSZ and
the EoR maps, the size of the ionized bubbles also play a key
role. Recall that the average size of the ionization bubble is larger
for `QSOs'. As a result, the underlying structure within the
ionized bubble will additionally reduce the anti-correlation and might
change the scale of (anti-)correlation.

\begin{figure}
\centering \includegraphics[width=.45\textwidth]{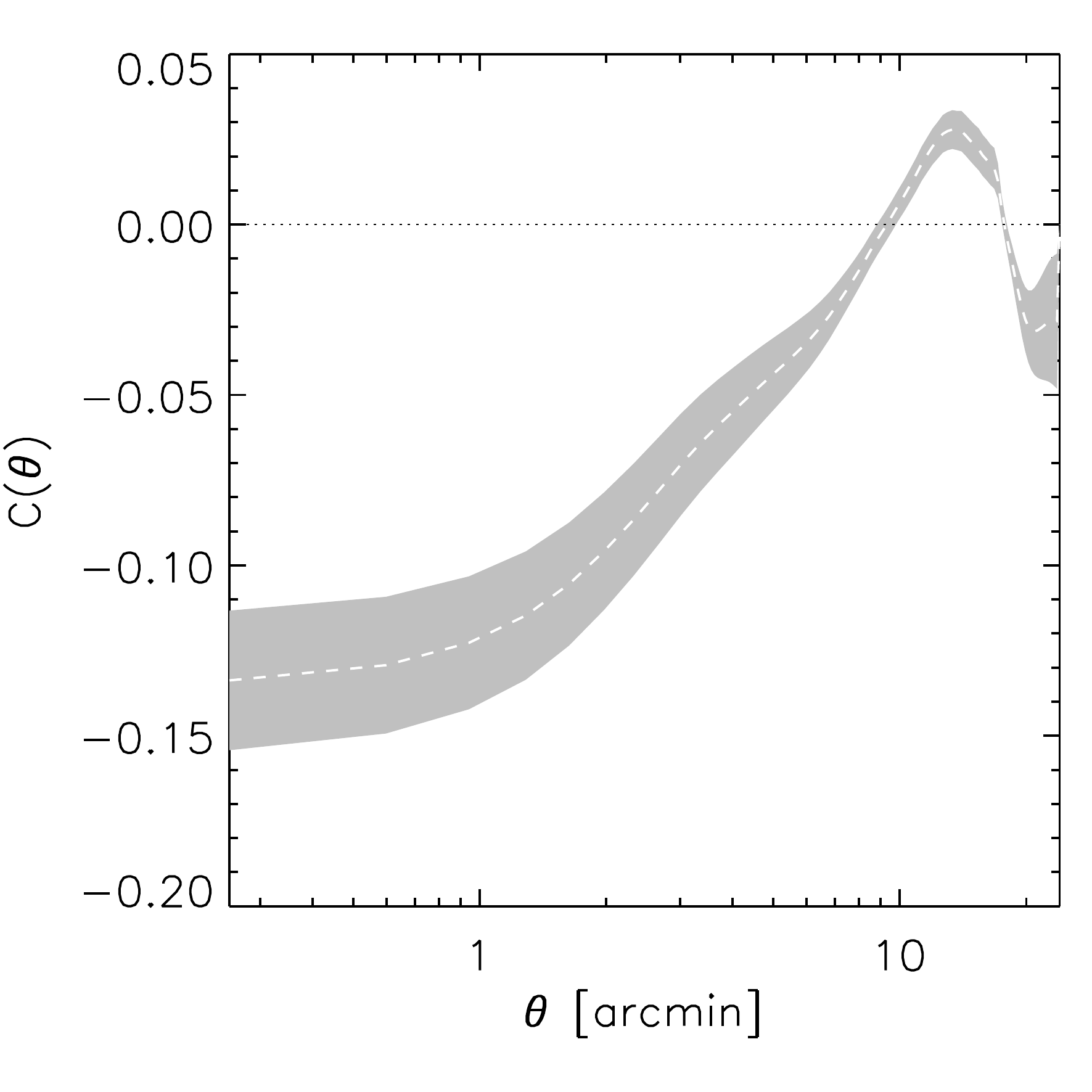}
\caption{\emph{The cross-correlation between integrated EoR and
integrated kSZ map as a function of lag ($C(\theta)$) for the `Stars'
reionization history scenario (dashed white line). The gray shaded
surface represents the estimated error obtained by Monte Carlo
simulation. Note that the correlation coefficient at the zero lag is
$C_{0}=-0.16\pm0.02$.}}
\label{fig:corrlag}
\end{figure}

From now on we will just concentrate on cross-correlations using
'Stars' since the `QSOs' model does not show a significant
anti-correlation. Fig.~\ref{fig:corrlag} shows the correlation
coefficient as a function of lag ($C(\theta)$) between the integrated
kSZ and the integrated EoR map. The dashed red lines represents the
estimated error obtained from Monte Carlo simulations. As in
\citet{salvaterra05}, we find that the two signals are anti-correlated
below a characteristic angular scale $\theta_c$ and this scale
indicates the average size of the ionized bubble which in our case
is $\theta_c \approx 10~{\rm arcmin}$.

\begin{figure}
\centering \includegraphics[width=.45\textwidth]{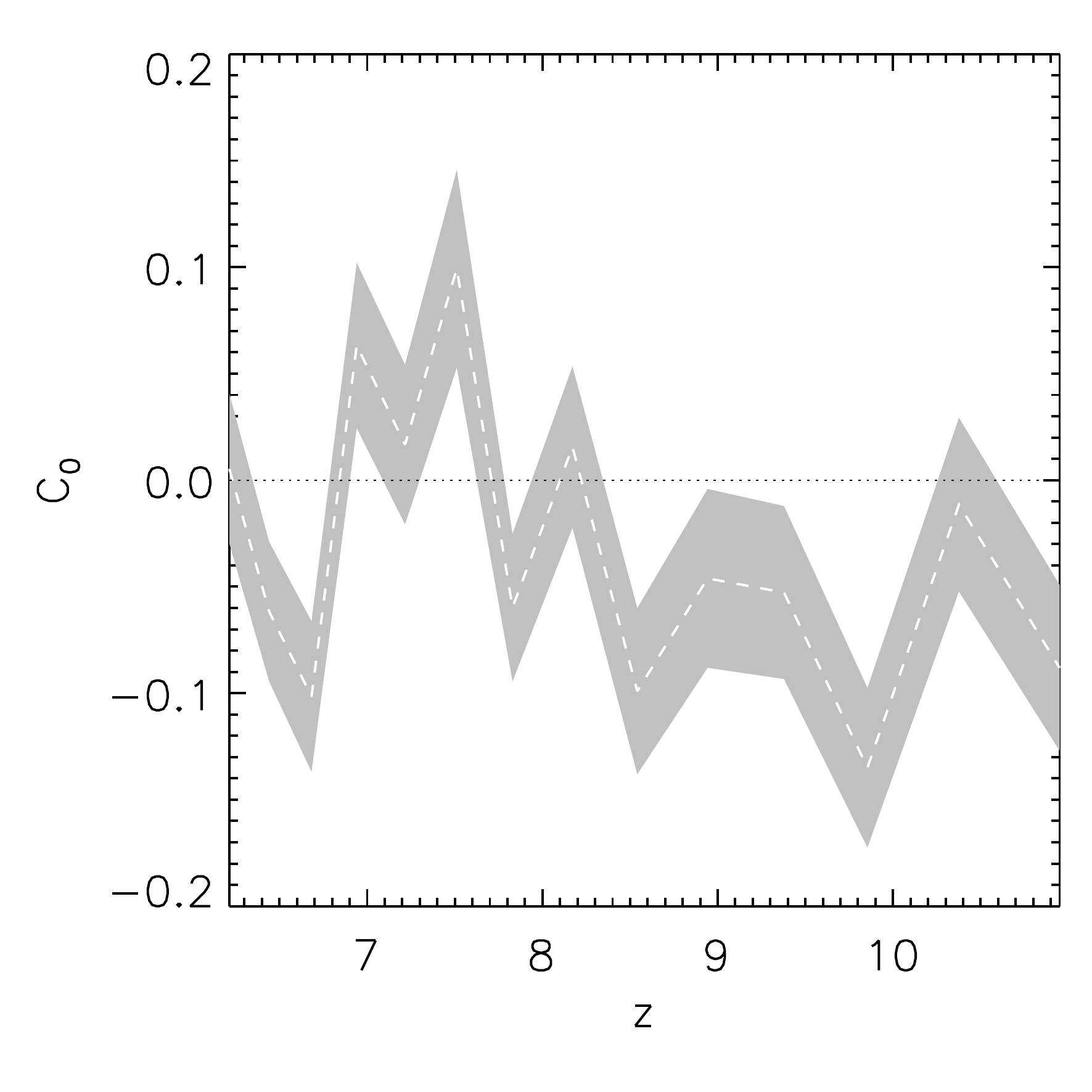}
\caption{\emph{The redshift evolution of the zero leg correlation
coefficient between the integrated kSZ map and the EoR map at the
certain redshift. The result is shown for the `Stars' reionization
history model. Note that the EoR map at the certain redshift is
produced by integrating $100~h^{-1}~{\rm Mpc}$ volume around the
desired redshift.}}
\label{fig:corr0EoRz}
\end{figure}

\citet{salvaterra05} also showed that the amplitude of the
anti-correlation signal increases with decreasing redshift and that
the characteristic angular scale shows a redshift evolution. In order
to test this in our simulation, we calculate the redshift evolution of
the zero lag cross-correlation coefficient between the integrated kSZ
map and the EoR map at different redshifts (Fig.~\ref{fig:corr0EoRz}).
To calculate the error in the cross-correlation, we generate 200
different realizations of the kSZ and corresponding EoR cubes using
the randomization procedure explained in Section~\ref{sec:sim}. Then,
around a desired redshift we fix the kSZ effect to zero and integrate
along non-zero part of the kSZ cube. Finally, we cross-correlate the
integrated kSZ map with the EoR map at the desired redshift and
estimate the error on the cross-correlation between the integrated kSZ
map and the EoR map at the certain redshift.

\begin{figure*}
\centering \includegraphics[width=1.\textwidth]{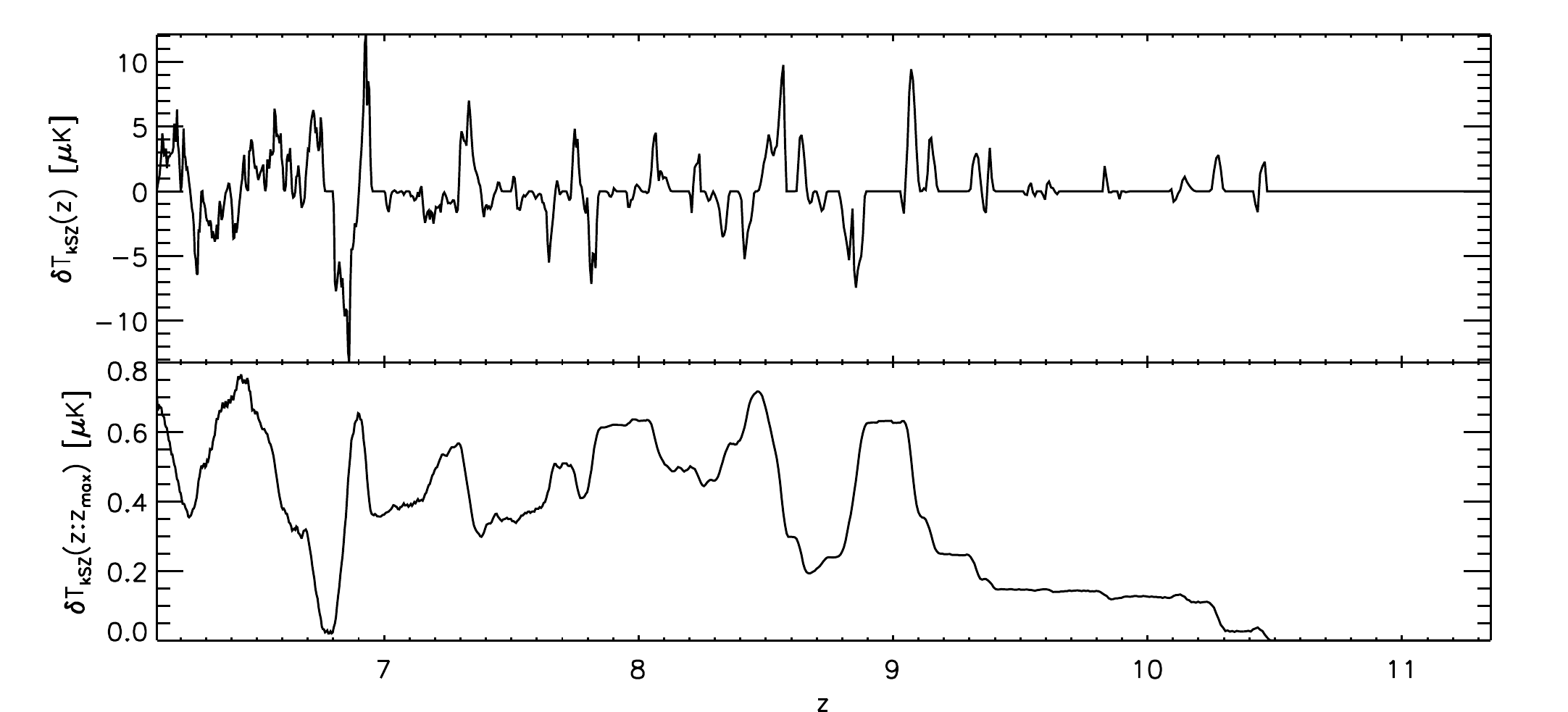}
\caption{\emph{\textsc{Top panel:} A random line of sight through
`Stars' kSZ cube, $\delta T_{\rm kSZ}(z)$, that is averaged over 10 pixels
($\sim 0.7~{\rm arcmin}$) at each redshift. \textsc{Bottom panel:} For
the same LOS the cumulative integral of the kSZ effect,$\int_{z_{\rm
max}}^z \delta T_{\rm kSZ}(z) {\rm d} z$ . Note that there is no coherent
redshift contribution to the integrated kSZ effect.}}
\label{fig:cumkSZz}
\end{figure*}

From Fig.~\ref{fig:corr0EoRz}, we find no coherent redshift evolution
of the anti-correlation signal and on the contrary at few redshifts
the two signals correlate instead of anti-correlating. The correlation
at a given redshift is caused by i) the patchy nature of the EoR
signal, which implies that there are some redshifts at which the EoR
map contains none or only a few small ionized bubbles. If one
correlates such an EoR map with the integrated kSZ map, the outcome is
a correlation between the two, and because of an insignificant number
of the ionization bubbles there is no contribution to the
anti-correlation. ii) the kSZ signal being patchy. There are some
redshifts where the kSZ signal from a certain ionization bubble does
not contribute significantly or at all to the integrated kSZ map (see
Fig.~\ref{fig:cumkSZz}). This could happen due to a weak kSZ signal
from a certain ionized bubble or due to cancellation of the kSZ signal
from another ionization bubble along the LOS.

The analysis is repeated for different binnings in frequency of the
EoR map but the results does not differ significantly. We also
calculate the redshift evolution of the charateristic angular scale
($\theta_c$), but we do not find any coherent evolution. This result
is driven by the fact that the contribution of the kSZ signal from a
certain redshift to the integrated kSZ map is not significant or is
even non-existent. As a result, if there is no coherent redshift
contribution to the integrated kSZ map there will be no coherent
redshift evolution of the kSZ-EoR cross-correlation signal (see
Fig.~\ref{fig:cumkSZz}).

The discrepancy between our results and that of \citet{salvaterra05}
is mainly due to the difference in the method to calculate the
cross-correlation coefficient. In Salvaterra et al., they first
calculate the cross-correlation coefficient (not normalized) between a
certain kSZ and EoR map. Then, they scramble both maps without keeping
any structural information and calculate the cross-correlation
coefficient. They compare the coefficients in the two cases to draw
their conclusion. In contrast to Salvaterra et al., we first calculate
the normalized cross-correlation coefficient (see Eq.~\ref{eq:c0})
between a pair of kSZ-EoR map. And then for comparison, we preform a
Monte Carlo simulation to generate different realizations of the kSZ
and the EoR maps . However, despite the cross-correlation procedure
used, once the primary CMB fluctuations are included we are not able
to find any significant kSZ-EoR cross-correlation (see
Sec.~\ref{sec:pCMB}).

\subsection{Large-scale velocity}
Our simulation volume is $(100~h^{-1}~{\rm Mpc})^3$ (see
Section~\ref{sec:sim}). Thus, large-scale velocities associated with
bulk motions, on scales $\gtrsim 100~h^{-1}~{\rm Mpc}$ are
missing. The missing velocities represent $\sim 50\%$ of the total
power in the velocity field as given by the linear theory.

\citet{iliev07} showed that the large-scale velocities on scales
$\gtrsim 100~h^{-1}~{\rm Mpc}$ increases the kSZ signal. Motivated by
this result, we approximately account for the missing large-scale
velocities as follows: first, we assume that every $100~h^{-1}~{\rm
Mpc}$ chunk of our simulation cube has a random large-scale
velocity component $v_{\rm LS}$. Since our simulation cube is produced
using 15 simulation boxes ($100~h^{-1}~{\rm Mpc}$), we need in total
15 $v_{\rm LS}$. We randomly choose a realization of the 15 $v_{\rm
LS}$ based on a velocity field power spectra from linear theory. By
doing this we ensure that the velocities are correlated at large
scales. Finally we add the missing $v_{\rm LS}$ component to each
$100~h^{-1}~{\rm Mpc}$ chunk of the simulated cube.

Based on 200 realizations of the large-scale velocity field, we have
found that the large-scale velocities increase the kSZ signal during
the EoR by 10\%. But on average we do not find any significant
increase or decrease in the kSZ-EoR cross-correlation. However, for
$\sim$20\% of all large-scale velocity realizations we find an
increase in the cross-correlation signal by a factor two or larger and
for $\sim$2\% a factor three or larger.

\subsection{Primary CMB}\label{sec:pCMB}
Up to now, in our cross-correlation analysis we only considered
secondary CMB anisotropies generated by the kSZ effect. In the actual
experiment, the CMB data will comprise not only the kSZ anisotropies
which are secondary, but also the primary and other secondary CMB
anisotropies (for a recent review see \citet{aghanim08}). In this
subsection we will examine the influence of the primary CMB
fluctuations on the detectability of the kSZ-EoR cross-correlation.

\begin{figure}
\centering \includegraphics[width=.45\textwidth]{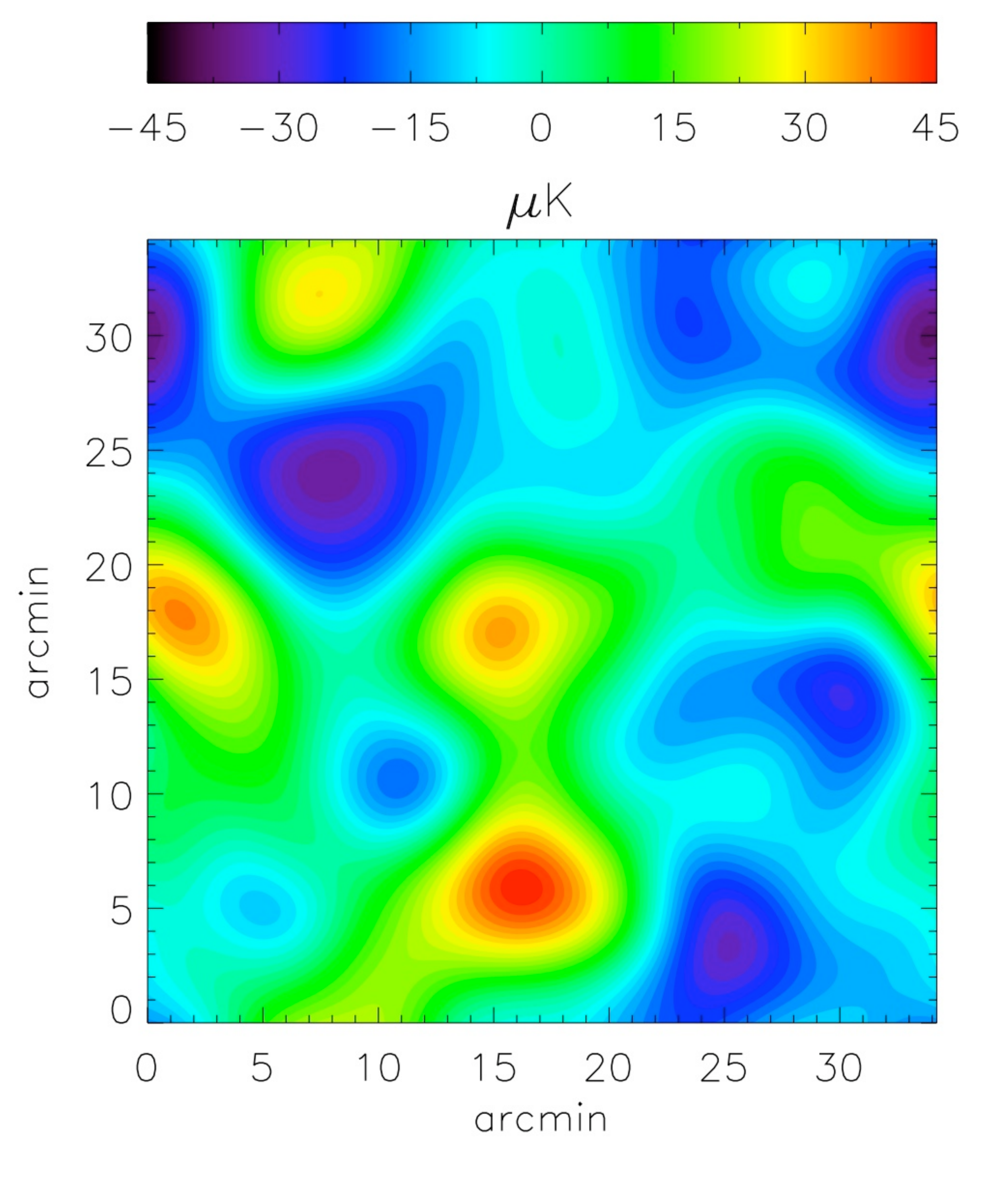}
\caption{\emph{The map of the primary CMB fluctuations generated as a
Gaussian random filed with a power spectrum obtained from the CMBFAST
algorithm.}}
\label{fig:pCMB}
\end{figure}

We simulate the primary CMB fluctuations in the following way: first
the CMB power spectra is obtained using CMBFAST (U. Seljak \&
M. Zaldarriaga, 2003) and then the map of the primary anisotropy is
produced as a random Gaussian field with this power spectrum.  An
example of the simulated primary CMB map is shown on the
Fig.~\ref{fig:pCMB}. The size of the map coresponds to the size of the
simulated EoR and kSZ maps. Notice the leak in power at small scales
due to Silk damping \citep{silk67}.

\begin{figure}
\centering \includegraphics[width=.45\textwidth]{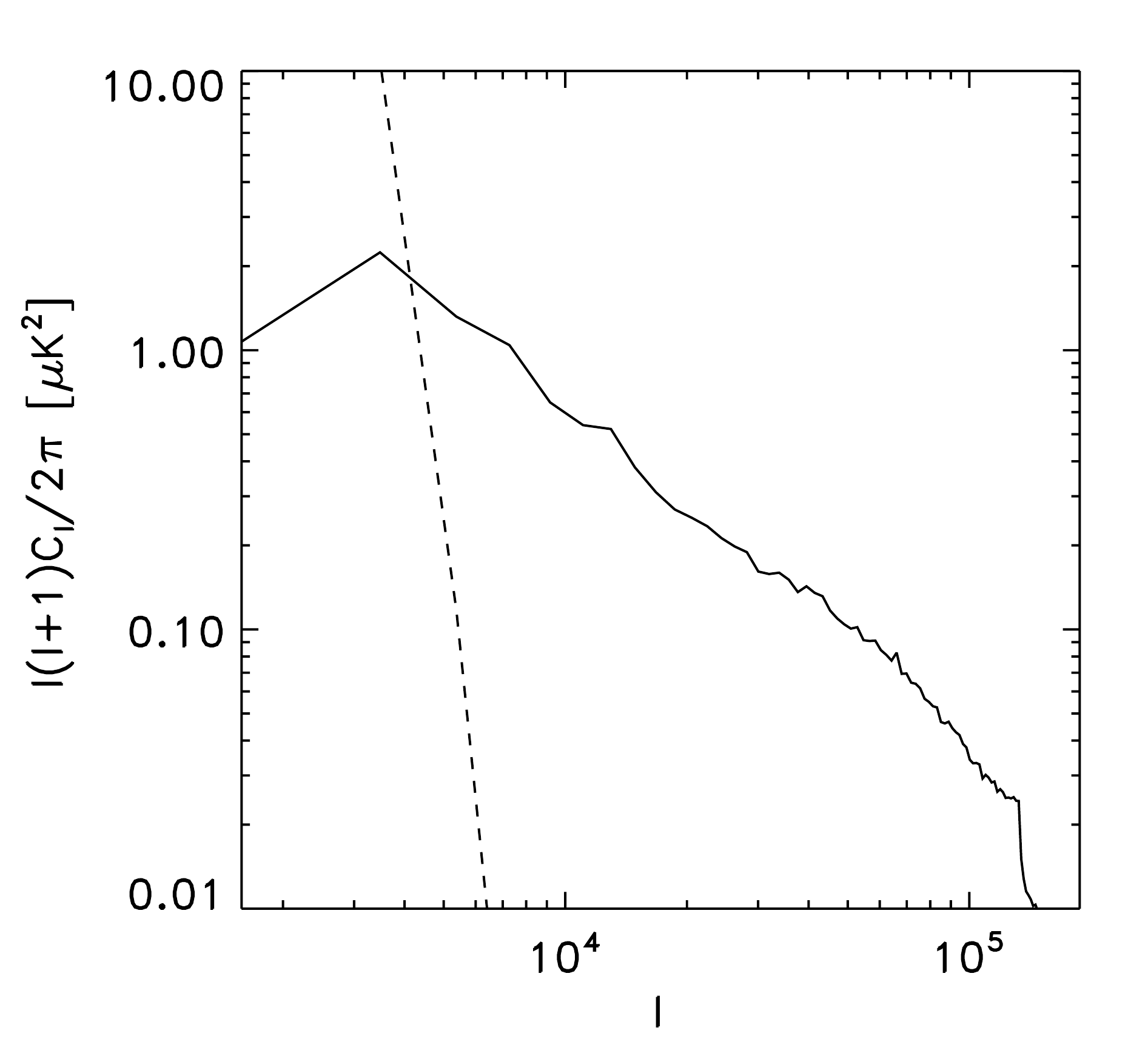}
\caption{\emph{The power spectra of primary CMB fluctuations (dotted line)
and kSZ anisotropies obtained from the simulated maps (solid line).}}
\label{fig:powspec}
\end{figure}

In order to calculate the noise in the cross-correlation introduced by
the primary CMB fluctuations, we generate 200 different realizations
of the primary CMB (pCMB) fluctuations. We then add secondary kSZ
anisotropies induced by the `Stars' (map shown in the
Fig.~\ref{fig:EoRkSZinteg}), and calculate the cross-correlation
between the pCMB+kSZ map and the integrated cosmological 21~cm map.
The obtained zero lag cross-correlation coefficient is
$0.0\pm0.3$. The noise introduced by the primary CMB fluctuations is
too large to find any significant kSZ-EoR (anti-)correlation.
However, one has to remember that the primary CMB anisotropies are
damped on small angular scales and that on these scales the secondary
anisotropies are the dominant component of the CMB power spectra (see
Fig.~\ref{fig:powspec}). Utilizing this fact, one can do a cross power
spectrum and see the correlation as a function of angular
scale. Pursuing this lead, we calculate the kSZ-EoR cross spectrum
first without and then with the primary CMB added to the kSZ map.

The cross spectrum ($C_{l}^{X}$) between the two images of a small
angular size is give by:
\begin{equation}\label{eq:crossspec}
C_{l}^{\rm X} \simeq P_{k}^{\rm X}=\frac{1}{n_k}\sum_{p,q\in k}
A_{p,q}\cdot B_{p,q}^{*},
\end{equation}
where $A_{p,q}$ is Fourier transform of the first image, $B_{p,q}^*$
the complex conjugate of the Fourier transform of the second image and
$n_k$ is number of points in the $k$-$\rm th$ bin ($k=\sqrt{p^2+q^2}$). 
Note that we assume a `flat-sky' approximation \citep[e.g.][]{white99}:
$k^2P(k)\simeq\frac{l(l+1)}{(2\pi)^2}C_l\mid_{l=2\pi k}$ which is
valid for $l\gtrsim60$.

\begin{figure}
\centering \includegraphics[width=.45\textwidth]{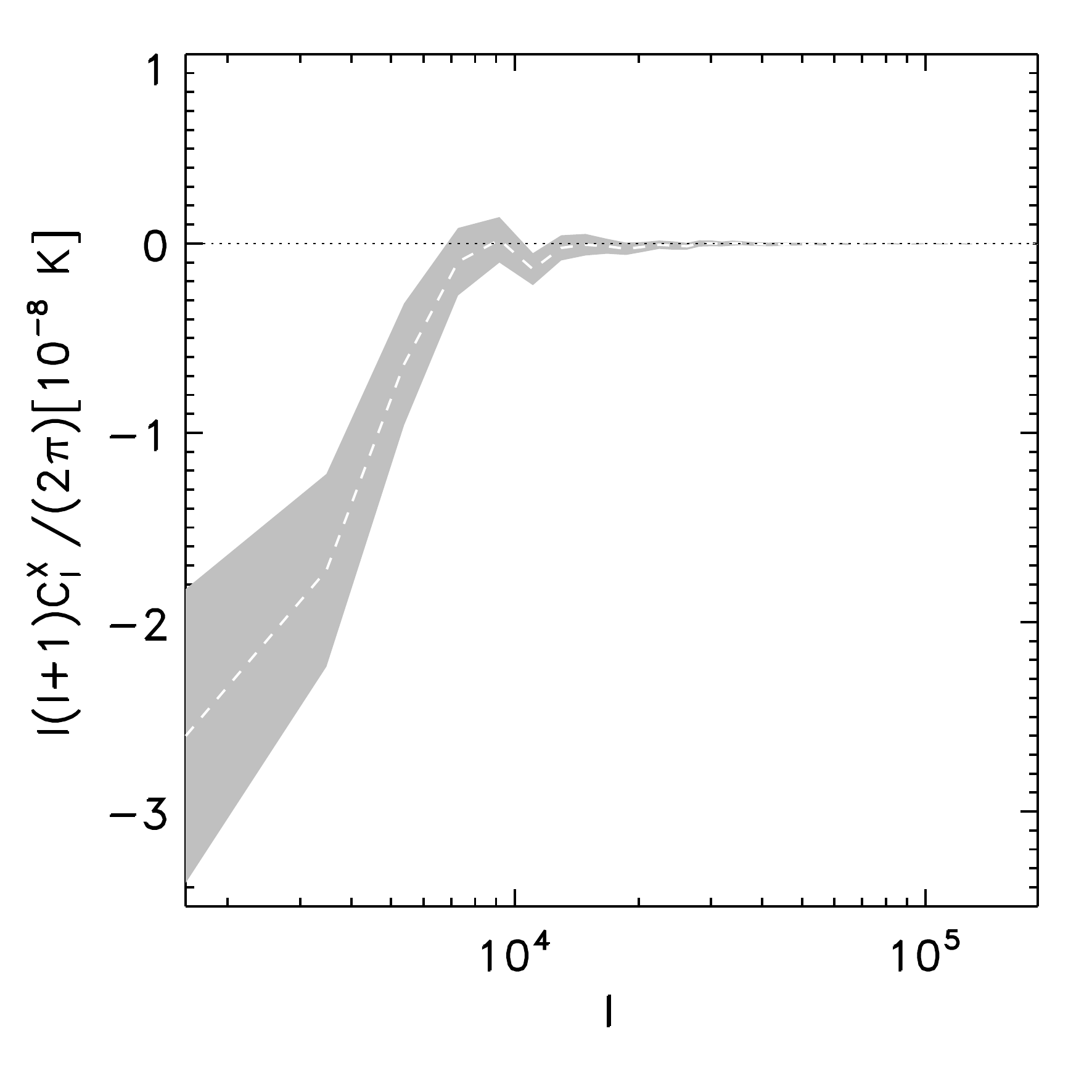}
\caption{\emph{The cross spectrum (see Eq.~\ref{eq:crossspec}) between
integrated kSZ map and integrated EoR map for the `Stars' reionization
history (dashed white line). The gray shaded
surface represents the estimated error obtained by Monte Carlo
simulation. Note that the primary CMB fluctuations are not included.}}
\label{fig:corrFS}
\end{figure}

Fig.~\ref{fig:corrFS} shows the cross power spectrum between the kSZ
anisotropies and the integrated cosmological 21~cm map for
reionization due to `Stars'. It is evident from the plot that the two
images anti-correlate on large scales ($l \lesssim 8000$) but that the
anti-correlation becomes weaker towards the smaller angular scales. At
angular scales $l \gtrsim 8000$, there is no (anti-)correlation.

We also calculate the cross power spectrum between integrated EoR map
and integrated kSZ map with primary CMB fluctuations
included. However, the noise introduced by the primary CMB is too
large to find any significant correlation at scales $l \lesssim 8000$.

This result might be driven by the simulation box size and
reionization scenarios considered in this study and does not mean that
a cross-correlation signal is absent at all scales and reionization
histories. In order to test this, one needs to explore the kSZ-EoR
cross-correlation using simulation with the box size that is larger
than the one used in this study ($100~h^{-1}~{\rm Mpc}$).

In the following subsection we try to use different cross-correlation methods 
and filtering techniques to explore the kSZ-EoR cross-correlation at scales
 where kSZ signal dominates over primary CMB fuctuations and where correlation 
signal is non-zero ($4000 \lesssim l \lesssim 8000$).

\subsection{Additional cross-correlation techniques}
For better understanding of the properties of the kSZ-EoR
cross-correlation, and with the hope of being able to find the
cross-correlation signal in the presence of primary CMB fluctuation,
in this subsection we apply techniques of filtering, wavelet
decomposition and relative entropy on our data. We will only use the
integrated kSZ map and integrated EoR map from the `Stars' model of
reionization, since the strongest cross-correlation signal is obtained
for this model in the previous sections. Note that in a following
analysis we first use only the kSZ and the EoR map and then as a
second step we include the primary CMB fluctuations.

\begin{figure*}
\centering \includegraphics[width=.33\textwidth]{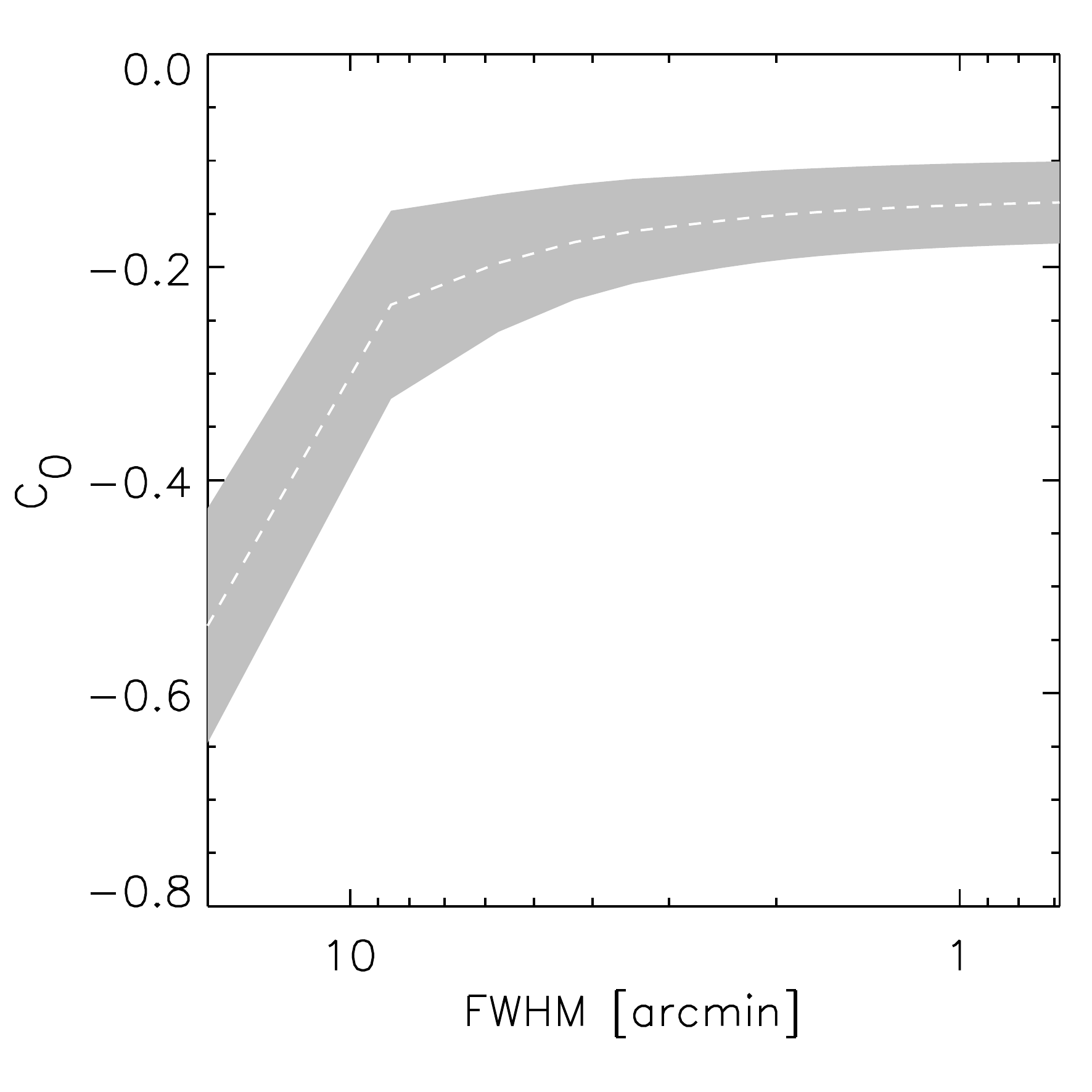}
\centering \includegraphics[width=.33\textwidth]{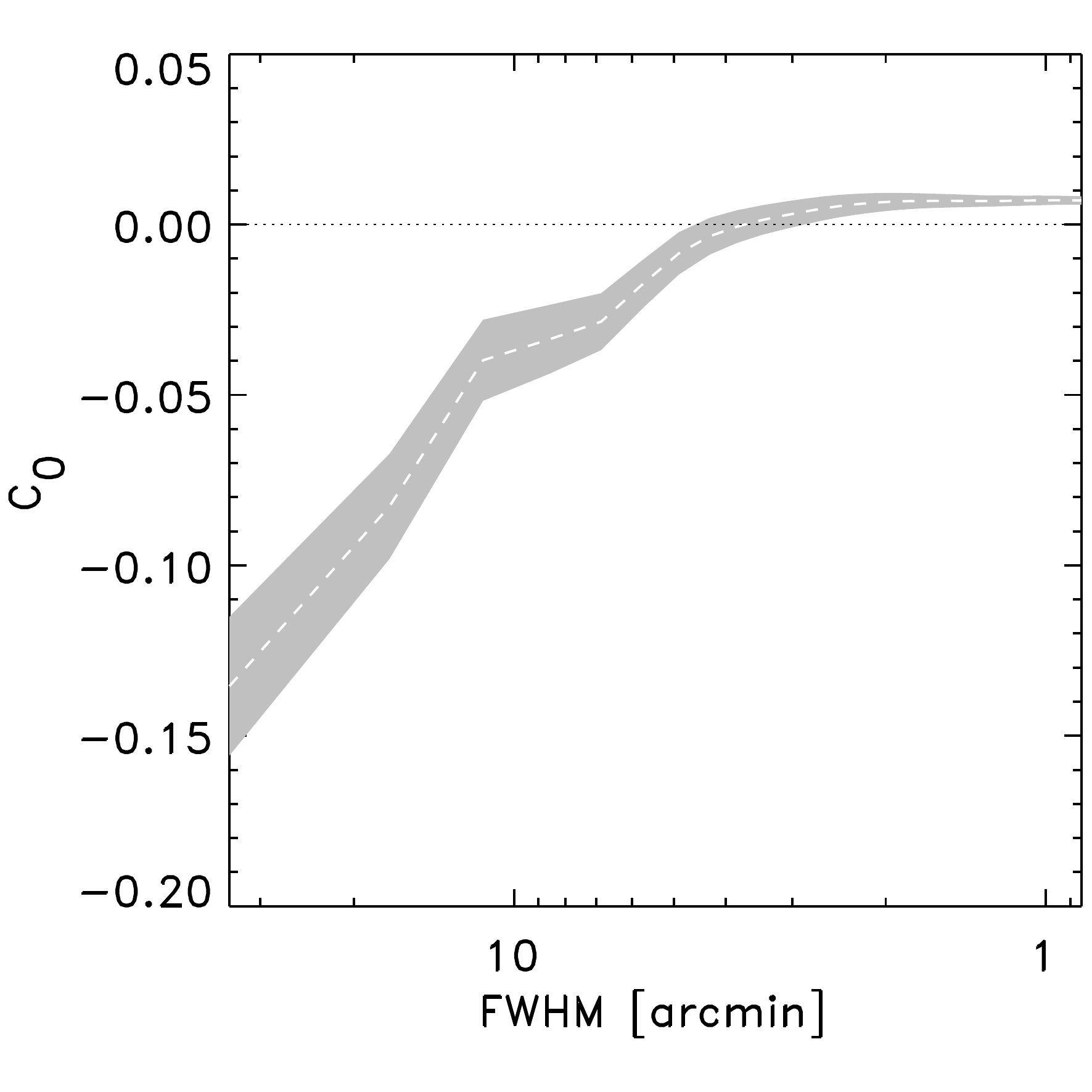}
\centering \includegraphics[width=.33\textwidth]{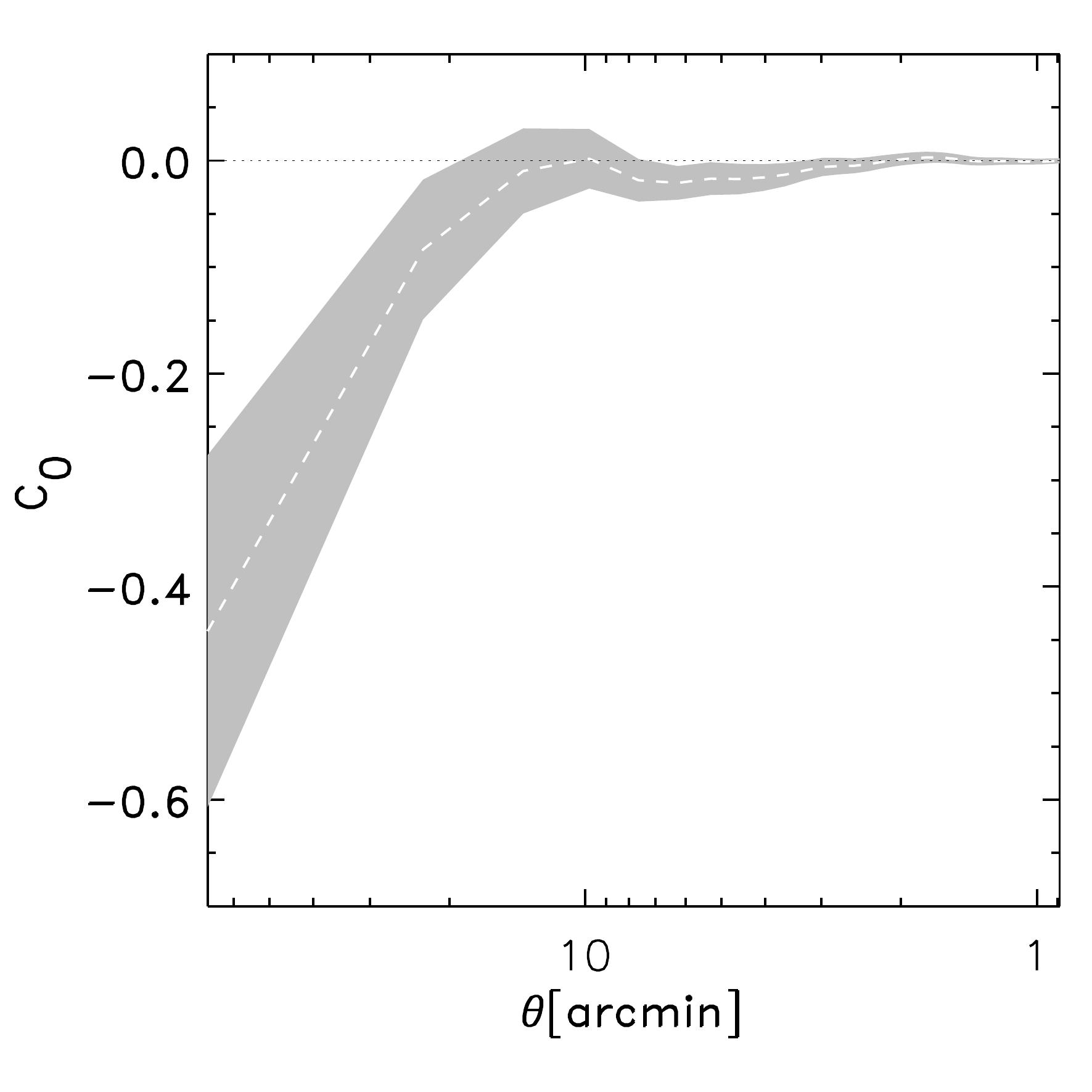}
\caption{\emph{The zero lag cross-correlation coefficient as a
function of the three different filtering procedures.  The first one
uses a high-pass filter, the second one uses low-pass filter and the
third a band-pass filter that passes only a certain scale.  In all
three cases the filter is based on the `Top hat' function. The dashed
white line is mean and the gray shaded surface represents the
estimated error obtained by Monte Carlo simulation.}}
\label{fig:TMHfilt}
\end{figure*}

Fig.~\ref{fig:TMHfilt} shows the zero lag cross-correlation
coefficient as a function of the three different filtering procedures.
The first one uses a high-pass, the second a low-pass and the third a
band-pass filter that passes out only a certain scale.  In all three
cases the filter is based on the `Top hat' function.  We filter out
the desired scale from both the kSZ map and the EoR map and calculate
the cross-correlation coefficient at zero lag. The results are shown
for the low-pass and high-pass filter as a function of the FWHM of the
filter and for the band-pass filter as a function of scale.

The plot on the left panel in Fig.~\ref{fig:TMHfilt} implies that the
anti-correlation is strongest on the largest scales of the map. By
adding smaller scales, the correlation coefficient decreases meaning
that smaller scales introduce noise in the correlation. The middle
panel in Fig.~\ref{fig:TMHfilt} suggests the same behaviour. By
removing the large scales, the cross-correlation signal becomes quite
small. Finally, the third panel of Fig.~\ref{fig:TMHfilt},
suggests that the large scales are indeed the dominant component of
the anti-correlation signal.

As a next step in our analysis we include the primary CMB
fluctuations. However we obtain the same result as discussed in the
previous subsection. On the scales where the kSZ anisotropies dominate
over primary, or the anti-correlation signal is too weak or the noise
introduced by residuals of the primary CMB fluctuations is too large
to find any statistically significant kSZ-EoR (anti-)correlation.

The wavelet analysis of the maps is done using Daubechies and Coiflet
wavelet functions. Both integrated kSZ map with added primary CMB
flucutations and integrated EoR map are decomposed to a certain
wavelet mode and then they are cross-correlated. Because the outcome
is similar to that of filtering we will not discuss this further.

The last method applied to the data is the ``relative entropy'', also
know as Kullback-Leibler distance. The relative entropy is a measure
of the shared information between two variables (two images) by
comparing the normalized distribution of the two. This method too
did not produce any significant result.

\section{Discussion and conclusions}\label{sec:dc}
This paper presents a cross-correlation study between the kinetic
Sunyaev-Zel'dovich (kSZ) effect and cosmological 21~cm signal produced
during the epoch of reionization (EoR). The study uses an N-body/SPH
simulation along with a 1-D radiative transfer code (the
\textsc{BEARS} algorithm, \citet{thomas09}) to simulate the EoR and to
obtain maps of the cosmological 21~cm signal and of the kSZ
effect. The maps are produced using the $100~h^{-1}~{\rm Mpc}$
co-moving simulation box for five different (3 homogeneous and 2
patchy) models of reionization history. The homogeneous model with
varying degree of ``rapidness'' of the reionization process is given
by Eq.~\ref{eq:hhist}. The patchy reionization histories include one
by `Stars' (gradual) and the other by `QSOs' (instant).

For a homogeneous reionization history we find that the kSZ
map and the integrated EoR map are correlated. Furthermore, that the
correlation depends on duration of reionization with larger values for
more ``rapid'' models. This result agrees with the analytical kSZ-EoR
cross-correlation analysis carried out by \citet{alvarez06}.

For patchy reionization models we find that the kSZ temperature
fluctuations are of the few $\mu K$ level (see Table~\ref{tab:rmsSQ}),
and is in agreement with previously obtained results by
\citet{salvaterra05, iliev07}. In addition, we show that the
temperature fluctuations induced by the patchiness of the reionization
process (`$\delta_{x_{\rm e}}$' term in Eq.~\ref{eq:dTkCMBz}) is
larger than the density induced fluctuations (homogeneous `$1+\delta$'
term in Eq.~\ref{eq:dTkCMBz}). The difference between the two is
stronger for the extended history (`Stars' model) than in more rapid
reionization histories (`QSOs' model) (see Table~\ref{tab:rmsSQ}).

As a first step in the kSZ-EoR cross-correlation study of patchy
reionization histories we cross-correlate the kSZ map and EoR map at
each redshift (see Fig.~\ref{fig:EoRkSZs}~\&~\ref{fig:EoRkSZq}). As
expected, the kSZ and the EoR map anti-correlate at certain redshifts
(see Fig.~\ref{fig:corr0z}).

We then cross-correlated the integrated cosmological 21~cm map and the
integrated kSZ map for patchy reionization (see
Fig.~\ref{fig:EoRkSZinteg}). The two signals show significant
anti-correlation only in the `Stars' model ($C_{0,{\rm
Stars}}=-0.16\pm0.02$, $C_{0,{\rm QSOs}}=-0.05\pm0.02$.). The result
is driven by the balance between homogeneous and patchy (`$1+\delta$'
and `$\delta_{x_{\rm e}}$' term in Eq.~\ref{eq:dTkCMBz}) kSZ
anisotropies and the average size of the ionized bubbles. Since the
homogeneous kSZ anisotropies correlate and patchy kSZ anisotropies
anti-correlate with the cosmological 21~cm maps, the two effects tend
to cancel each other. In addition the average size of the ionization
bubble is larger for `QSOs' than in `Stars' model and the structure of
matter within the ionized bubble reduce the cross-correlation. As a
consequence the kSZ-EoR anti-correlation is much stronger for the
extended (`Stars' model) reionization history than for a more instant
history (`QSOs' model).

For a patchy model of reionization we estimated the redshift evolution
of the correlation coefficient ($C_0$) and characteristic angular
scale $\theta_{\rm C}$. This was done by cross-correlating the
integrated kSZ maps with the EoR maps at different redshifts (see
Fig.~\ref{fig:corr0EoRz}). In contrast to \citet{salvaterra05}, we do
not find any significant coherent redshift evolution of $C_0$ and
$\theta_{\rm C}$. The discrepancy between the results is caused by the
difference in a procedure used for calculating cross-correlation.
However, despite the cross-correlation procedure used, once the
primary CMB fluctuations are included we are not able to find any
significant kSZ-EoR cross-correlation.

The influence of the missing large-scale velocities on the kSZ signal
and kSZ-EoR cross-correlation was investigated. Although, the
large-scale velocities increase the kSZ signal by 10\%, we do not
find, on average, any significant change in the kSZ-EoR
cross-correlation. However, for $\sim$20\% of large-scale velocity
realizations we find an increase in the cross-correlation signal by a
factor two or larger and for $\sim$2\% a factor three or larger.

Data from CMB experiments contains both the secondary (e.g. kSZ) and
primary anisotropies. For completeness of our study we calculated the
noise in the kSZ-EoR cross-correlation introduced by the primary CMB
fluctuations and found that its addition reduces the cross-correlation
signal to zero ($C_0=0.0\pm0.3$). The cross-correlation was also
performed on scales where the kSZ anisotropies dominate over the
primary CMB fluctuations ($l \gtrsim 4000$, see
Fig.~\ref{fig:powspec}). We have done this by calculating cross-power
spectra (Fig.~\ref{fig:corrFS}), applying different filtering methods
(Fig.~\ref{fig:TMHfilt}) on the data and by doing wavelet
decomposition. However, the outcome of the analysis is that on the
scales where the kSZ anisotropies dominate over primary, or the
anti-correlation signal is too weak or the noise introduced by
residuals of the primary CMB fluctuations is too large to find any
statistically significant kSZ-EoR (anti-)correlation.

As a sanity check we calculate the kSZ-EoR cross-correlation
using the simulation obtained by \citet{iliev07} (`f250C'
$100~h^{-1}~{\rm Mpc}$ simulation). The reionization history of this
model is similar to our `QSOs' model. The reionization history is
relatively sharp and instant. The cross-correlation coefficient at
zero lag for the integrated kSZ map and integrated EoR map is
$C_0=-0.04\pm0.02$. The result is in agreement with the result
obtained from the `QSOs' model. We also calculated the redshift
evolution of the zero lag cross-correlation coefficient and have found
no coherent redshift evolution.

In view of all the results obtained from the kSZ-EoR cross-correlation
study, we conclude that the kSZ-EoR anti-correlation on scales
captured by our simulation box ($\sim 0.6^\circ $) is not a reliable
technique for probing the EoR. However, there is still hope that we
will be able to find the correlation between the kSZ and EoR signal on
 scales larger than $\sim 1^\circ $, where the patchiness of the
ionization bubbles should average out (see \citet{alvarez06} and
Tashiro et al., \textit{in preparation}). Finally, it is important to
note that the kSZ signal induced during the EoR could still be
detected in the power spectra of the CMB and used to place some
additional constrains on this epoch in the history of our Universe.

\section*{acknowledgement}
We acknowledge discussion with the LOFAR EoR key project members.  The
authors thank I. Iliev for providing us with \citet{iliev07} (`f250C'
$100~h^{-1}~{\rm Mpc}$) simulation, J. Schaye and A. Pawlik for their dark matter 
simulation and A. Ferrara for useful comments.
As LOFAR members VJ, SZ, LVEK and RMT are partly funded by the
European Union, European Regional Development Fund, and by
`Samenwerkingsverband Noord-Nederland', EZ/KOMPAS.

\bibliographystyle{mn2e}
\bibliography{reflist}

\appendix

\bsp

\label{lastpage}

\end{document}